\documentclass[modern]{aastex61}

\usepackage{graphicx}
\usepackage{amssymb}
\usepackage{amsmath}

\usepackage{natbib}
\DeclareMathOperator*{\argmax}{argmax}

\definecolor{awesome}{rgb}{1.0, 0.13, 0.32}

\received{}
\revised{}
\accepted{}

\shorttitle{Fast algorithms for slow moving asteroids}
\shortauthors{Whidden et al.}

\begin{document}

\title{Fast algorithms for slow moving asteroids: constraints on the distribution of Kuiper Belt Objects}

\correspondingauthor{J. Bryce Kalmbach}
\email{brycek@uw.edu}

\author{Peter J. Whidden} \affil{Department of Astronomy, University of Washington, Seattle, WA, 98195, USA}
\author{J. Bryce Kalmbach} \affil{Department of Physics, University of Washington, Seattle, WA, 98195, USA}
\author{Andrew J. Connolly} \affil{Department of Astronomy, University of Washington, Seattle, WA, 98195, USA}
\author{R. Lynne Jones} \affil{Department of Astronomy, University of Washington, Seattle, WA, 98195, USA}
\author{Hayden Smotherman} \affil{Department of Astronomy, University of Washington, Seattle, WA, 98195, USA}
\author{Dino Bektesevic} \affil{Department of Astronomy, University of Washington, Seattle, WA, 98195, USA}
\author{Colin Slater} \affil{Department of Astronomy, University of Washington, Seattle, WA, 98195, USA}
\author{Andrew C. Becker} \affil{Amazon Web Services, Seattle, WA, 98121, USA}

\author{\v{Z}eljko Ivezi\'c} \affil{Department of Astronomy, University of Washington, Seattle, WA, 98195, USA}

\author{Mario Juri\'c} \affil{Department of Astronomy, University of Washington, Seattle, WA, 98195, USA}
\author{Bryce Bolin} \affil{Department of Astronomy, University of Washington, Seattle, WA, 98195, USA}
\author{Joachim Moeyens} \affil{Department of Astronomy, University of Washington, Seattle, WA, 98195, USA}
\author{Francisco F\"orster} \affil{Center for Mathematical Modeling, Beaucheff 851, 7th floor, Santiago, Chile} \affil{Millennium Institute of Astrophysics, Chile}
\author{V. Zach Golkhou} \affil{Department of Astronomy, University of Washington, Seattle, WA, 98195, USA} \affil{The eScience Institute, University of Washington, Seattle, WA 98195, USA}



\begin{abstract}
We introduce a new computational technique for searching for faint moving sources in astronomical images. Starting from a maximum likelihood estimate for the probability of the detection of a source within a series of images, we develop a massively parallel algorithm for searching through candidate asteroid trajectories that utilizes Graphics Processing Units (GPU). This technique can search over $10^{10}$ possible asteroid trajectories in stacks of the order 10-15 4K x 4K images in under a minute using a single consumer grade GPU. We apply this algorithm to data from the 2015 campaign of the High Cadence Transient Survey (HiTS) obtained with the Dark Energy Camera (DECam). We find 39 previously unknown Kuiper Belt Objects in the 150 square degrees of the survey. Comparing these asteroids to an existing model for the inclination distribution of the Kuiper Belt we demonstrate that we recover a KBO population above our detection limit consistent with previous studies. Software used in this analysis is made available as an open source package.
\end{abstract}



\section{Introduction}

Traditional approaches for detecting Trans Neptunian Objects (TNOs) rely on the identification of sources within individual images and then linking these sources to generate orbits \citep{kubica+2007,denneau+2013}. More recently digital tracking or ``shift-and-stack" techniques have been developed to search for moving sources below the detection limit of any individual image \citep{Gladman+1997, Allen+2001,Bernstein+2004,Heinze+2015}. These approaches are fundamentally different from the traditional techniques in that they assume a trajectory for an asteroid and align a set of individual images along that trajectory in order to look for evidence for a source. 

Shift-and-stack methods share many commonalities with the ``track before detect''
(TBD) method used for the tracking of satellites and
missiles \citep[e.g.][]{reed+1988, johnston+2002}. This field is mature in
literature and implementation, and has been generalized to enable the
detection of not just linear motion, but also the tracking of
``acutely maneuvering non--cooperative
targets'' \citep{rozovskii+1999}.  We will adopt several
of the features of TBD, in particular those described
in \cite{johnston+2002}, who outline the core principles of accumulating the
track detection probability, and in quantifying the false alarm
probability.  

A related approach to faint moving object detection is
presented in \cite{Lang+2009}, who describe a search for
high proper motion stars.  These objects move by a distance comparable
to the PSF FWHM over the course of an entire survey (i.e. years).
Thus a direct image stack is sufficient to detect objects.
However, \cite{Lang+2009} return to the individual science
images to perform a joint fit for the proper motion and parallax of
the objects, even though they appear at low signal--to--noise in any
individual image.  The scaling of detection depth in these techniques
goes formally as $\Delta m$ = 1.26 log($N$) magnitudes, or 1 magnitude
deeper after the linking of 6 faint detections, and 2 magnitudes after
40 detections.  

The advantage of digital tracking is that we increase the detection limit for a series of $N$ images as $\sqrt{N}$ (assuming a constant point spread function (PSF) and background across all images). For objects having power--law distributions in apparent magnitude -- e.g. the double power--law TNO model of \cite{Bernstein+2004} -- linking 6 epochs of data would yield an increase of 4--7 times as many objects {\it from the same data}. The cost of digital tracking comes from the combinatorial and computational complexity of having to search a large number of candidate trajectories for each pixel within an image.
Digital tracking must combine the individual images along a proposed motion vector that will depend on the assumed distance of the asteroid. Even for slow moving asteroids searches will scale as $Nn$ where $n$ is the number of pixels in an image.

These computational costs have limited the application of digital tracking to searches for slowly moving objects or to narrow ``pencil-beam'' surveys. In this paper we introduce a new approach that utilizes a probabilistic formalism for the detection of sources in images (removing the need to stack or coadd images) and  Graphics Processing Units (GPUs) to massively parallelize the number of searches that can be undertaken concurrently. In \S 2 we introduce a maximum-likelihood formalism for the detection of sources and extend this for the case of moving objects. In \S 3 we apply this approach to the High Cadence Transient Survey (HiTS) and describe some of the filtering techniques that were applied to exclude false positives in the candidate asteroids. In \S 4 we discuss the asteroids detected by this approach and compare their properties to current models and observations.

\section{Fast tracking and stacking of images}

\begin{figure}
	\centering
    	\includegraphics[width=\linewidth]{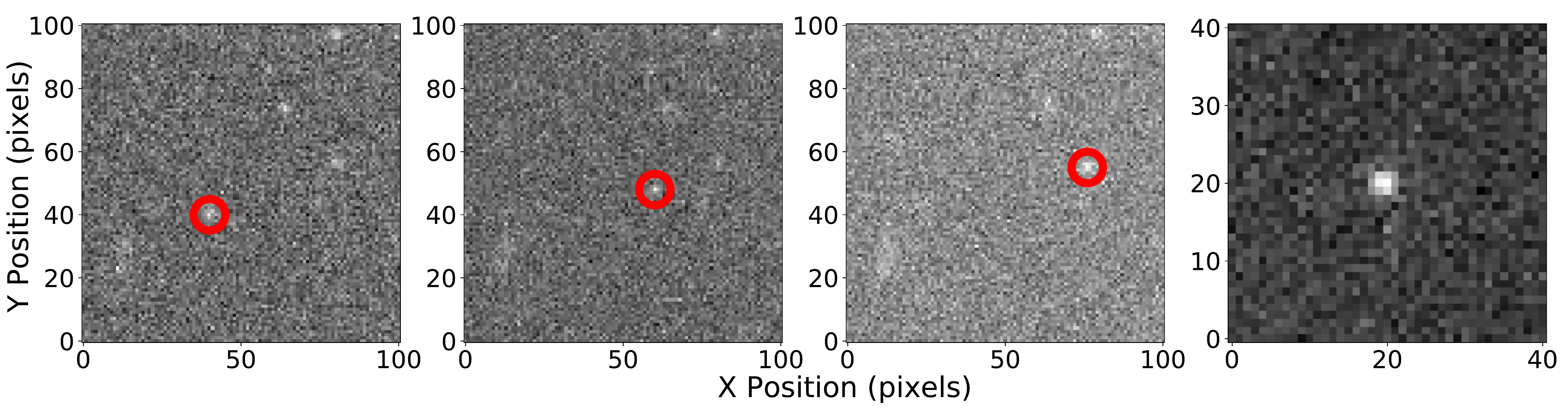}
	\caption{Shifting and stacking of individual images along the asteroid's trajectory creates a single point source in the stacked image.}	\label{shift_stack_demo}
\end{figure}

Digital tracking assumes a set of $N$ images have been observed over a period of time (from minutes to days) with the individual images covering approximately the same part of the sky. The individual images are astrometrically shifted along a proposed
motion vector, coadded, and then searched for point sources in the resulting coadds \citep{Gladman+1997,Allen+2001,Bernstein+2004}. This approach is illustrated in Figure \ref{shift_stack_demo} where the image on the far right is the sum of the previous three images added along the motion of the highlighted object. Creating a stack optimized for faint, moving objects must include astrometric
offsets between the images that correspond to the distance that an object has 
moved between observations. Unfortunately, this angular velocity
is not known \textit{a priori}, and in fact differs for each moving object in
the field.  Thus a stacked search for Solar System objects in
time--series data requires a sequence of coadds, each optimized for a
particular motion vector.
Due to the combinatorial complexity
of the problem, these efforts have traditionally been optimized for
TNO recovery, where apparent angular motions are small.

During an observation of a Solar System object, there are apparent
motion contributions from the object's own space velocity, as well as
from the reflex motion of the Earth, which is primarily due to the
Earth's motion around the Sun but also includes the Earth's rotation
around its axis.  These contribute to an apparent angular motion of
the Solar System object, which will trail during an
exposure.  If the object trails by more than the PSF full-width half-maximum (FWHM) during
observation, its signal is spread over additional background pixels,
which lessens the overall signal-to-noise \citep{Shao+2014}.

These ``trailing losses" also apply to a
stacked image: the image stack velocity must be close enough to the
true velocity of the object to not spread the signal over more than
the PSF FWHM.  This requirement, together with the range of expected
apparent motions of the desired objects, sets the number or sampling of the velocities
(or orbits) that must be searched and stacks that must be examined.
Typical apparent motions (at opposition) range from $20\arcsec$/hr for
main belt asteroids at 3 AU to $1\arcsec$/hr for TNOs.

\subsection{A likelihood-based approach for source detection} \label{sec:likelihood_detection}

\subsubsection{Form of single pixel likelihood function} \label{singlePix}

Our goal is to derive the likelihood functions for a given source
being present in the coaddition of a series of images.  First we start
by finding the form of the likelihood for a single pixel in a single
image. The following derivation is based upon work in \citet{kaiser} and interpreted in \citet{bosch} and \citet{Bosch+2017}.
Photons landing on a pixel in a detector follow Poisson
statistics. For a given pixel the probability of counting
$n$ photons with an expected value of $\mu$ is defined as
\begin{equation} \label{eq:1}
P(n|\mu)= \frac{\mu^{n}e^{-\mu}}{n!}.
\end{equation}
This probability can also be interpreted as the likelihood that a
model of the sources within an image (hereafter the true image)
generates a predicted count $\mu$ given we observe $n$ counts on a
pixel. The log-likelihood of our prediction model is, therefore,
\begin{equation}\label{eq:2}
\mathcal{L}(\text{model}) = \ln P(\text{data}|\text{model}) = \ln P(n|\mu) = n \ln \mu - \mu - \ln n!
\end{equation}
Differentiating $\mathcal{L}$ with respect to $\mu$, the
log-likelihood has its peak when $\mu_{o} = n$. Furthermore, if we
Taylor expand around this maximum value then we get
\begin{equation}\label{eq:3}
\mathcal{L}(\text{model}) = n \ln \mu_{o} - \mu_{o} - \ln n! + (\frac{n}{\mu_{o}} - 1)(\mu - \mu_{o}) - \frac{n}{2\mu_{o}^{2}}(\mu - \mu_{o})^{2} + \ldots
\end{equation}
\begin{equation}\label{eq:4}
\mathcal{L}(\text{model}) = \text{constant} - \frac{1}{2} \frac{(n - \mu)^{2}}{n} + \ldots
\end{equation}
where the constant contains all of the terms that depend only on $n$ and so
are independent of our model. Finally, if $n$ is large we can ignore
higher order terms and approximate our likelihood function as that of
a Gaussian with likelihood proportional to
$e^{- \frac{(n - \mu)^{2}}{2n}}$.

\subsubsection{Pixels and PSF}
At this point we need to take a step back and understand what exactly
goes into calculating photon counts at each pixel. To do this we will
follow the derivation laid out in \citet{bosch} but for a
two-dimensional image. First, we start with the number of counts $n(x, y)$ we get
from the true sky intensity $I(x, y)$ on a pixel centered at
$(x,y)$. The observation on our detector will be the true sky
intensity convolved with the PSF, $T(x, y)$. In addition, there will
be an extra integral to account for the binning into a pixel centered
at $(x,y)$ with side length $a$. That gives us
\begin{equation}
n(x,y) = \int_{x-a/2}^{x+a/2} \int_{y-a/2}^{y+a/2} dv \ du \int_{-\infty}^{\infty} \int_{-\infty}^{\infty} dq \ dp \ I(p,q)\ T(u-p, v-q),
\end{equation}
which can be rewritten in terms of two convolutions where we rewrite the pixel binning integral as a convolution with a square top hat function $H(x,y)$ with height 1 and side length $a$
\begin{gather}
n(x, y) = \int_{-\infty}^{\infty} \int_{-\infty}^{\infty} dv \ du \ H(x-u, y-v) \int_{-\infty}^{\infty} \int_{-\infty}^{\infty}  dq \ dp \ I(p,q)\ T(u-p, v-q) \\
n(x,y) = [H \ast I \ast T](x,y)
\end{gather}
and if we exploit the associative and commutative properties of convolutions we can rewrite this as 
\begin{equation}
n(x,y) = [I \ast (H \ast T)](x,y)
\end{equation}
and following the procedure in \citet{bosch} simplifying the term in parentheses to
\begin{equation}
T_{a}(x,y) = (H \ast T)(x,y)
\end{equation}
and finally ending up with
\begin{equation}
n(x,y) = [I \ast T_{a}](x,y)
\end{equation}
so that when we refer to the PSF below as $T_{a}$ we are referring to
the PSF including the pixel transfer function.

\subsubsection{Full likelihood function for a single image}
Now we understand how to represent the pixel data as a function of the sky and PSF as well as how to write out the likelihood function of a single pixel. We consider an image $\textbf{x}$, where the $i$th pixel is $\textbf{x}_{i}$ and the real sky is modeled as $f(\textbf{x})$. Then we use our assumption of a large photon count to justify a Gaussian likelihood function as described in \ref{singlePix} and get
\begin{equation}
L(\textbf{x}_i) = P(\text{data}|\text{model}) = P(n(\textbf{x}_{i})|f(\textbf{x}_{i})) = \frac{1}{\sqrt{2\pi\sigma_{\textbf{x}_{i}}^{2}}}e^\frac{-\frac{1}{2}(n(\textbf{x}_{i})-[f \ast T_{a}](\textbf{x}_{i}))^{2}}{\sigma_{\textbf{x}_{i}}^2}
\end{equation}
but if we are looking at the entire image plane we need to go from a single variable Gaussian to a multivariate Gaussian distribution. Thus, the likelihood for the full image is
\begin{equation}
L(\textbf{x}) = P(n(\textbf{x})|f(\textbf{x})) = \frac{1}{\sqrt{(2\pi)^{k}|C|}} e^{-\frac{1}{2}\sum_{i,j} (n(\textbf{x}_{i})-[f \ast T_{a}](\textbf{x}_{i})) \times C^{-1}(\textbf{x}_{i}, \textbf{x}_{j}) \times(n(\textbf{x}_{j})-[f \ast T_{a}](\textbf{x}_{j}))}
\end{equation}
where $k$ is the number of pixels and $C$ is the pixel covariance matrix. If we take the log-likelihood function then we have 
\begin{equation}\label{fullLikeForm}
\mathcal{L}(\textbf{x}) = -\frac{k}{2}\ln(2\pi) - \frac{1}{2}\ln|C| -\frac{1}{2}\sum_{i,j} (n(\textbf{x}_{i})-[f \ast T_{a}](\textbf{x}_{i})) \times C^{-1}(\textbf{x}_{i}, \textbf{x}_{j}) \times (n(\textbf{x}_{j})-[f \ast T_{a}](\textbf{x}_{j}))
\end{equation}
for the full form of the log-likelihood of a given sky model $f(\textbf{x})$. 

\subsubsection{Likelihood of detection of a point source}
If we want to find a point source at a pixel $\textbf{y}$ in the image with flux $\alpha$ then the sky model we are proposing is $f(\textbf{x}) = \alpha \delta({\textbf{x} - \textbf{y}})$, a delta function located at the pixel lcoation of the source multiplied by the flux. Now let $n(\textbf{x})$ represent the flux counts for each pixel we have in a background subtracted image and notice that the first two terms in equation \ref{fullLikeForm} are independent of the model $f(\textbf{x})$. Furthermore, in order to keep values positive, we change to the negative log-likelihood function which we are now trying to minimize in order to get the most likely parameters. Thus, we get 
\begin{equation}
\mathcal{L}(\alpha, \textbf{y}) = \text{constant} +\frac{1}{2}\sum_{i,j} (n(\textbf{x}_{i})-\alpha T(\textbf{x}_{i}-\textbf{y})) \times C^{-1}(\textbf{x}_{i}, \textbf{x}_{j}) \times (n(\textbf{x}_{j})-\alpha T(\textbf{x}_{j}-\textbf{y}))
\end{equation}
where the constant holds terms that are
model independent. Also, we have defined $\alpha T(\textbf{x}_{i}-\textbf{y}) = [f \ast T_{a}](\textbf{x}_{i})$ which is the convolution of the point source sky model with the PSF and is thus equivalent to the PSF centered at $\textbf{y}$.
$T(\textbf{x}_{j}-\textbf{y}))$ is the same for the pixels in the $j$ indexed sum. We have two sums here since we are summing across every combination of pixels from the covariance matrix. If we multiply out these terms it looks like
the following:
\begin{multline}
\mathcal{L}(\alpha, \textbf{y}) = \text{constant} + \frac{1}{2} \sum_{i,j} C^{-1}(\textbf{x}_{i}, \textbf{x}_{j})(n(\textbf{x}_{i}) n(\textbf{x}_{j})) - \alpha \sum_{i,j} C^{-1}(\textbf{x}_{i}, \textbf{x}_{j}) n(\textbf{x}_{i})T(\textbf{x}_{j} - \textbf{y}) \\
+ \frac{\alpha^{2}}{2} \sum_{i,j} C^{-1}(\textbf{x}_{i}, \textbf{x}_{j}) T(\textbf{x}_{i} - \textbf{y}) T(\textbf{x}_{j} - \textbf{y})
\end{multline}
Furthermore we define the following terms
\begin{gather}
\Psi(\textbf{y}) = \sum_{i,j} C^{-1}(\textbf{x}_{i}, \textbf{x}_{j}) n(\textbf{x}_{i})T(\textbf{x}_{j} - \textbf{y}) \\
\Phi(\textbf{y}) = \sum_{i,j} C^{-1}(\textbf{x}_{i}, \textbf{x}_{j}) T(\textbf{x}_{i} - \textbf{y}) T(\textbf{x}_{j} - \textbf{y})
\end{gather}
and once again add into constant the terms that are not model dependent. Finally, this gives us the likelihood in a compact form
\begin{equation}
\mathcal{L(\alpha, \textbf{y})} = \text{constant} - \alpha \Psi(\textbf{y}) + \frac{\alpha^{2}}{2} \Phi(\textbf{y})
\end{equation}
where $\Psi(\textbf{y})$ and $\Phi(\textbf{y})$ are new types of images that can be created using the PSF.


\subsubsection{Coaddition of likelihood images for point source detection} \label{sec:coadd_likelihood}

We make the assumption that the signal for the majority of candidate detections will be dominated by the background noise and that the background noise is independent in each pixel. While there may be sources of correlation between pixels such as electronic detector effects (e.g.\ crosstalk or the dependence of the PSF on intensity), for this treatment we will assume these effects are small enough to ignore and thus will continue as if we have a completely diagonal covariance matrix. This simplifies our calculations into images that can be precomputed with a simple kernel that approximates the PSF.

Thus, our equations for $\Psi$ and $\Phi$ images are reduced to 
\begin{gather}
\Psi(\textbf{y}) = \sum_{i} \frac{1}{\sigma_{i}^{2}} \ n(\textbf{x}_{i}) T(\textbf{x}_{i} - \textbf{y}) \\
\Phi(\textbf{y}) = \sum_{i} \frac{1}{\sigma_{i}^{2}} \ T(\textbf{x}_{i} - \textbf{y})^{2}
\end{gather}
where $\Psi$ is the inverse-variance weighted cross-correlation of the PSF and the data, which is also the convolution of the PSF rotated by 180 degrees or just the PSF if it is symmetric.  $\Phi$ is the effective area of the PSF weighted by the inverse variance \citep{Bosch+2017}.

Bright pixels in the image can be dealt with by a mask and we set the inverse variance to zero so that they will add nothing to the likelihood sum. Since trajectories will only cross over this pixel once when we run them over a series of images this serves to give noisy pixels zero weight in the full sum of a trajectory while keeping the information in the other images. 

Solving for the maximum-likelihood solution we end up with 
\begin{equation}
\frac{\partial \mathcal{L}_{ML}}{\partial \alpha} = - \Psi(\textbf{y}) + \alpha \Phi(\textbf{y}) = 0 \label{max_likelihood_derivative}
\end{equation}
and as a result we find that,
\begin{gather}
\alpha_{ML} = \frac{\Psi(\textbf{y})}{\Phi(\textbf{y})} \label{maxLikeFlux}
\end{gather}
and
\begin{gather}
\mathcal{L}_{ML}(\alpha_{ML}, \textbf{y}) = \text{constant} - \frac{\Psi^{2}(\textbf{y})}{2\Phi^{2}(\textbf{y})} \label{finalLikelihood}
\end{gather}
where $\alpha_{ML}$ is the most likely flux for a source at pixel $x_i$ and $\mathcal{L}_{ML}(\alpha_{ML}, \textbf{y})$ is the probability
of that source given the observation $n(x_i)$. 
The kernel that maximizes the likelihood of our flux measurements is, therefore, the 180 degree rotation of the PSF or, in the case of a symmetric PSF, the PSF itself.
Additionally,
Equation \ref{finalLikelihood} is the log form of a Gaussian---where by
analogy $\Psi \simeq (x-\mu)$ and $\Phi^{2} \simeq \sigma^{2}$---so we
can approximate a $\chi^{2}$ distribution. If we define $\nu =
\frac{\Psi}{\sqrt{\Phi}}$ and then make an image of $\nu$, we can call
points above some threshold value $m$ to be $m$-sigma detections \citep{szalay+1999}.

In order to make a coadded likelihood image, we create our $\Psi$ and $\Phi$ images from sums across all pixels in all images. This amounts to making $\Psi$ and $\Phi$ likelihood images for each of our original images, $i$, with the appropriate PSF for each respective image and summing them separately so that 
\begin{gather}
\Psi_{coadd} = \sum_{i} \Psi_{i}({\textbf{y}_{i}}) \\
\Phi_{coadd} = \sum_{i} \Phi_{i}({\textbf{y}_{i}}) \\
\nu_{coadd} = \frac{\Psi_{coadd}}{\sqrt{\Phi_{coadd}}}
\end{gather}
Here, $\nu$ is the signal-to-noise of the detection of a source at pixel $x$ within the coadded image. Therefore, the final value for the likelihood of a point object is just the sum of the values at a given set of points in the $\Psi$ image divided by the sum of the values at the same points in the $\Phi$ image. This means that if we wish to do a moving object detection, all that is needed is to sum the $\Psi$ and $\Phi$ values as shown, but we must use the appropriate pixel coordinates for a given trajectory as the $\textbf{y}_{i}$ values in each image---there is no need to shift and stack the likelihood images at any point. The $\Psi$ and $\Phi$ images can also be precalculated meaning we are able to store them in memory as many trajectories are searched.
    
\subsection{Object detection as Optimization}
Detecting moving point sources in a stack of images can now be approached as an optimization problem. For the linear trajectories we are considering, the core task can be reduced to finding the initial positions $\textbf{y}_{0}$ and velocities \textbf{v} so that $\nu$ is maximized. For a given potential object, the best candidate source is then
\begin{gather}
 \mathbf{\argmax}_{\textbf{y}_{0},\textbf{v}} (\nu_{coadd})
\end{gather}

\begin{figure}
	\centering
    	\includegraphics[width=\linewidth]{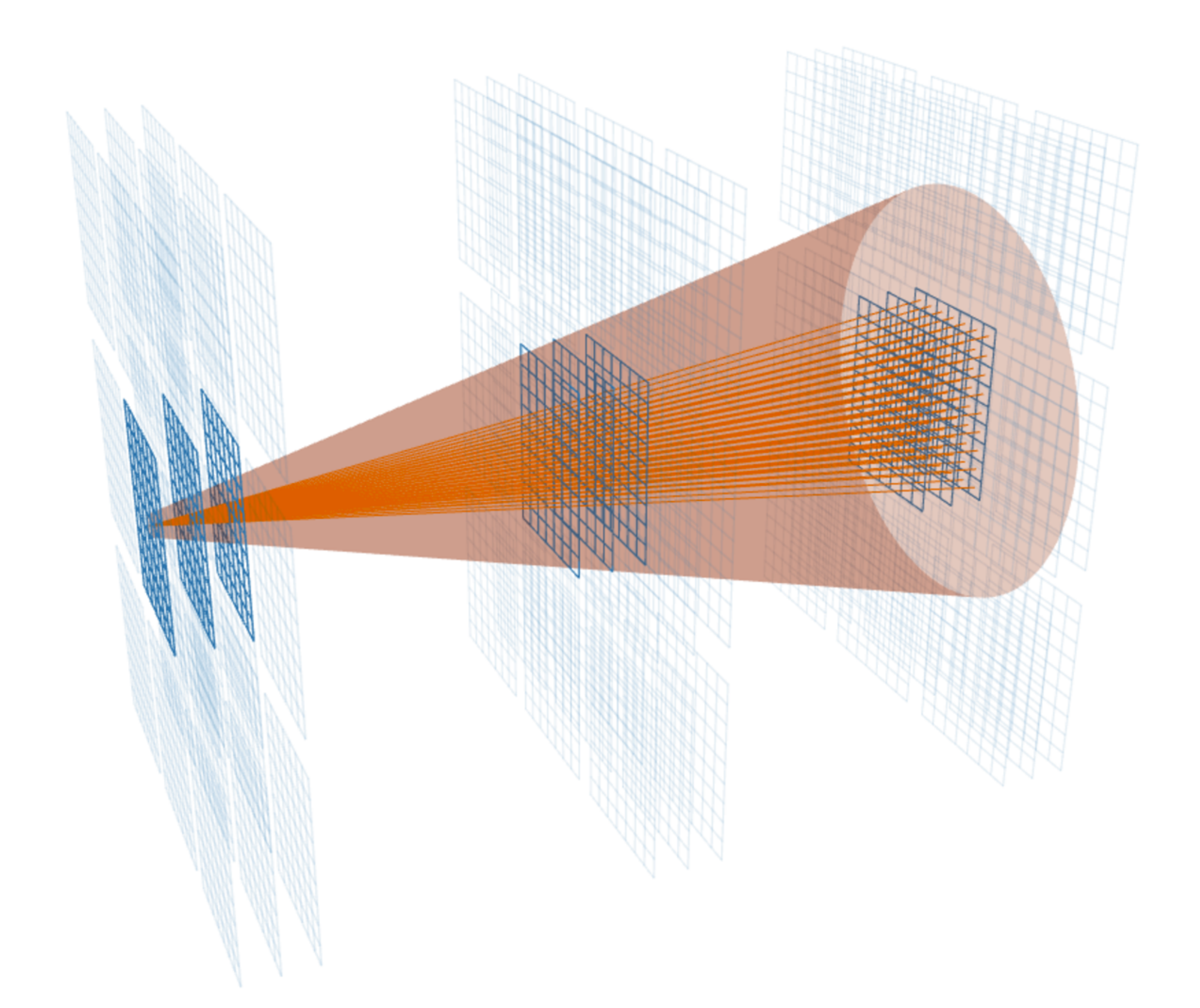}
	\caption{ Visualization of the many trajectories that must be searched in order to cover a defined velocity and angle range over a stack of images of the same field taken at different times.}	\label{searchtree}
\end{figure}

We hope to find all unique $\textbf{y}_{0}$ and \textbf{v} such that $\nu(\textbf{y}_{0}, \textbf{v})$ is above a desired detection threshold. The most straightforward computational approach to this problem is to evaluate $\nu$ for all possible values of $\textbf{y}_{0}$ and \textbf{v} that describe realistic orbits. An illustration of this approach is shown in Figure \ref{searchtree} starting from a single value of $\textbf{y}_{0}$. This method directly computes $\Psi$ and $\Phi$ for every relevant trajectory through all images by sampling pixels along each trajectory and storing the result if the integrated $\nu$ is above the threshold. The computational complexity of this algorithm is bounded by $\mathcal{O}(na(tu)^2/p)$ where $n$ is the number of images to be stacked, $a$ is the area on the sky to be searched, $p$ is the area of a single pixel, $t$ is the duration between the first and last image, and $u$ is the range of an object's apparent velocity. The complexity scales with time and velocity range quadratically because all trajectories ending anywhere inside an area with radius $tu$ must be considered. 

\subsection{GPU Implementation}
Comprehensively searching a stack of images for moving objects directly requires computing the value of $\nu$ billions of times. Fortunately each evaluation of $\nu$ is entirely independent from the others, and this allows for natural parallel execution. Rectangular patches of adjacent trajectories are grouped into thread blocks with dimensions 32x16 which are distributed across the GPU's multiprocessors. $\Psi$ and $\Phi$ pixels are interleaved in memory and within each thread block horizontally adjacent trajectories access them contiguously, enabling high throughput. To achieve good performance all images must be stored in the GPU's memory at once, which limits the size and number of images that can be used to about 100 4K x 4K images. In general, the runtime of this algorithm can be estimated as $Nnk$ where $N$ is the number of images in the stack, $n$ is the number of trajectories, and $k$ is a performance constant that is experimentally determined by the computer hardware and implementation. In our implementation we have a measured value of $k$ about $2.4\times10^{-11}$ seconds per image for each trajectory. Practically this means searching for a wide range of objects (240 billion trajectories) in a stack of 30 images (4K x 4K) takes about 180 seconds. 
Besides computing $\nu$, we also use the GPU for convolving the images. The convolution is done in a single pass with a spatially invariant kernel. In this work we only used a Gaussian PSF approximation, and could have used a two-pass separable convolution but chose to use a slower single pass to support non-Gaussian PSF's in the future.

\section{Searching for faint KBOS} \label{sec:search_description}

\subsection{The High Cadence Transient Survey (HiTS)}\label{sec:hits}

We tested our software using the first three nights (February 17-19, 2015) of the 2015 campaign of the High Cadence Transient Survey (HiTS) \citep{HITS}. The 2015 HiTS campaign involved repeatedly visiting 50 3-square-degree fields in the DECam \textit{g}-band filter. There were typically 5 visits per night to each of the 50 fields, taken with a 1.6hr cadence. The remaining three nights of the 2015 HiTS campaign data, also taken with the DECam \textit{g}-band, were used for follow up of the detected objects \citep{HITS}.

These data were taken using the Dark Energy Camera (DECam) at Cerro Tololo Interamerican Observatory (CTIO). The best quartile of seeing at CTIO is about 0.4'' full width at half maximum (FWHM). DECam has 60 2K x 4K CCDs, each with a pixel scale of 0.26 arcsec/pixel. We processed all of the data using the LSST Data Management (DM) Software \citep{lsst-dm} and ran our software using the warped science images that were laid out in 4K x 4K pixel patches. This means that the effective field of view of each warped image is about 0.34 square degrees  \citep{Flaugher_DECam_Instrument}. Because we currently do not search trajectories across CCDs, this is the effective field of view of any individual search. We also used the masks and variance planes from the LSST DM processing output.

\subsection{Application of the KBO search}

We created $\Psi$ and $\Phi$ images from the individual science images. To mask static objects we identify all sources that are detected at more than 5$\sigma$ above the background and mask the pixels associated with these sources. Masks for each static source are grown by an additional two pixels (in radius) to exclude lower surface brightness halos around these sources. To ensure that we do not mask bright \textit{moving} sources we require that any masked pixel in a science image must be masked in at least one of the other science images (i.e.\  a source must be present at the same position in two of the images for it to be defined as static and masked). We apply the union of the mask for all individual images as a global mask to the final science images.



Our search started at every pixel in the earliest observation (in time) for each HITS field and covered a range of 32,000 trajectories across the stacks of 12-13 images corresponding to three nights of HITS data. The 32,000 trajectories came from a linearly-spaced grid of 128 angular steps within $\pm 12^{\circ}$ of the ecliptic and 250 steps in velocities ranging from 1"-- 5.7"/hour. This grid was set up so that at the end of the search period, our maximum separation between the final pixels in a search pattern would be no more than approximately 2 PSF widths. This means that that we would be within 1 PSF width of any possible trajectory. For this search, we went down to a signal-to-noise ratio (SNR) threshold of 10, as measured by the $\nu$ parameter introduced in Section \ref{sec:coadd_likelihood}. For these data, this corresponds to a single image limiting magnitude in \textit{g} of 23.1.

    
\subsection{Filtering of candidate trajectories}
      
\begin{figure}
	\centering
    	\includegraphics[width=\linewidth]{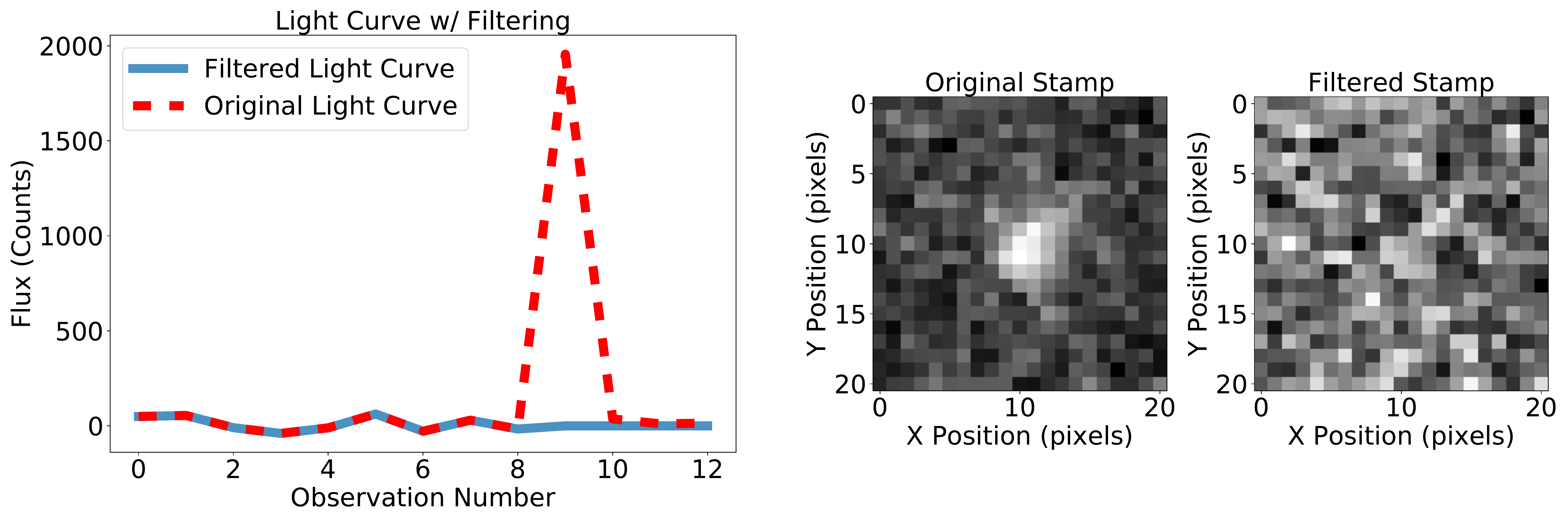}
	\caption{Left: Change in light curve when an image with an outlier flux is removed from the light curve. Right: Shifted and stacked postage stamps before and after outlier removal. After the single outlier observation is removed by the filter the trajectory is obviously not following a true object and is discarded.}	\label{filter_demo}
\end{figure}

After running the initial GPU search, we use a set of Python tools to filter out false detections. One such tool is an outlier filtering process that modifies estimates of the flux and variance of an object given a sequence of observations. This is done using the light curve of the candidate trajectory and a variation on the Kalman filter \citep{kalman} commonly used in signal processing. We filter out any observations that are more than five standard deviations away from the best fit Kalman flux at that observation and use the remaining observations to recalculate the likelihood of the candidate. As an example of this process, Figure \ref{filter_demo} shows postage stamps of the candidate before and after the filtering. In this case a fast, bright, moving object moved across the trajectory in a single image. Excluding that image leads us to the correct decision to discard this candidate trajectory, as the likelihood of an object along this trajectory drops to near zero, once the interloping object is removed.

\begin{figure}
	\centering
    	\includegraphics[width=\linewidth]{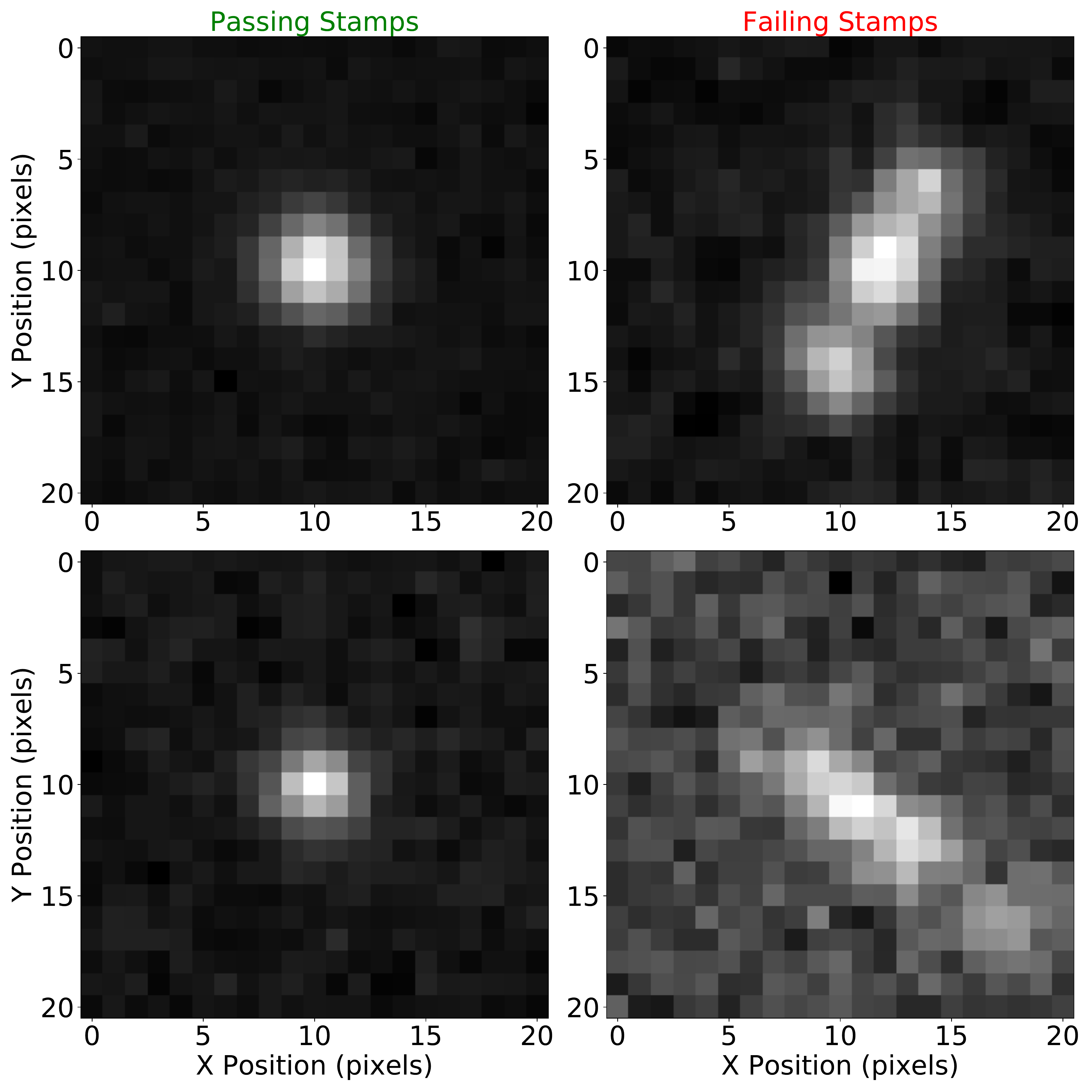}
	\caption{Left Column: Stamps that passed the lightcurve filter and the image moment filter. Right Column: Stamps that passed the lightcurve filter but were rejected by the image moment filter.}	\label{moment_filter}
\end{figure}

After the outlier detection we build coadded postage stamps for all remaining candidate trajectories. We then use scikit-image \citep{skimage} to calculate the central moments of the images and filter based upon the similarity of the moments to a Gaussian centered at the middle of the stamp. This does a good job of eliminating elongated shapes and trajectories where a bright source appeared in a single image but was off center. For example, in Figure \ref{moment_filter} the left column shows postage stamps of real objects that passed through the filtering. The right column of the figure shows postage stamps that made it through the outlier filtering but were ruled out after the image moment filter.

Our final step in the filtering process is clustering to remove duplicate results from the same object. We take the starting x and y pixels and the horizontal and vertical velocities of the candidate trajectories and use the DBSCAN clustering method \citep{dbscan} in the scikit-learn python package \citep{scikit} to group similar trajectories. We then take the highest likelihood trajectory for each group and save the results with postage stamps and light curves to file for final examination by eye.

\subsection{Found Objects} \label{subsec:found_objects}
\subsubsection{KBO properties in the HiTS field}


In total we found 45 KBOs in our search, of which only 6 were previously detected by the Pan-STARRS 1 (PS1) survey according to the Minor Planet Center (MPC). PS1 used the Pan-STARRS Moving Object Processing System (MOPS) which links detections from sources identified in individual difference images \citep{denneau+2013}.

\begin{figure}
	\centering
    	\includegraphics[width=\linewidth]{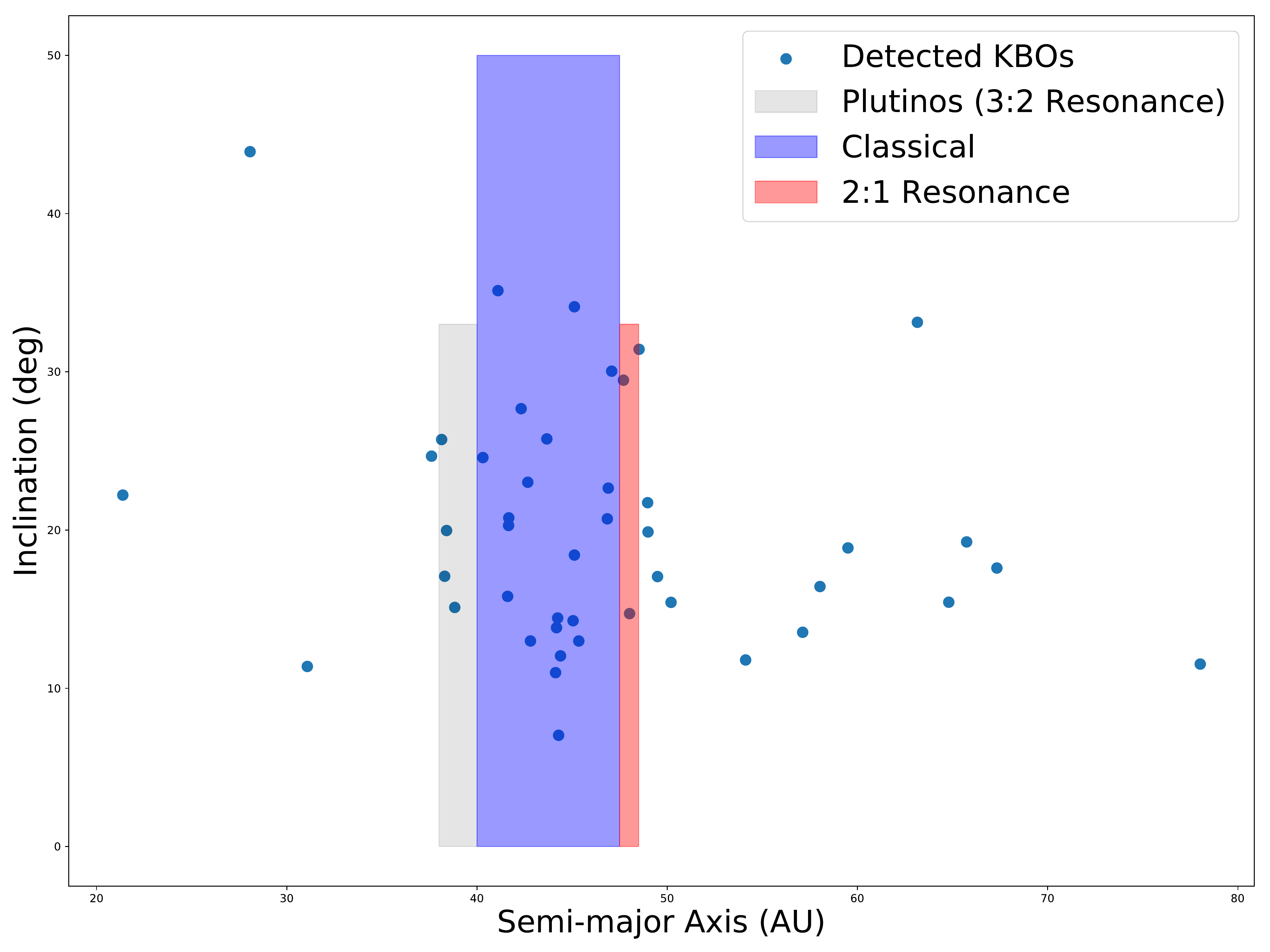}
	\caption{Semi-major axis versus inclination for detected objects with the range of different KBO populations highlighted.}	\label{semimajor_axis_inclination}
\end{figure}

The full information for all our object detections is shown in Table \ref{object_table} in Appendix \ref{sec:object_table}. We used the orbit fitting code of \citet{bernstein} for initial orbit determination and used the remaining HITS data (see Section~\ref{sec:hits}) to get additional observations where possible before submitting to the MPC. From this we estimated that, for the 45 KBOs, semi-major axes ranged from 21 AU to 67 AU and DECam \textit{g}-band magnitudes ranged from 22.1 to 24.7 mags. Figure \ref{semimajor_axis_inclination} compares semi-major axis to inclination for our discoveries and shows the overlap with KBO populations commonly discussed in the literature.

\subsubsection{Comparison with known asteroids} 
The candidate objects were checked with the MPC database of known objects. Of the 45 objects that were detected, 6 were known. Four of these objects were detected in our search down to a SNR threshold of 10 and they are noted in Table \ref{object_table}. The 2 remaining objects were much fainter at V magnitudes of 24.1 and 24.3 while the faintest MPC object we recovered in our search was at V = 23.0. We attempted to recover the objects at a fainter SNR threshold by rerunning the search on the image stacks corresponding to the predicted locations. One of the objects was recovered when we reduced our SNR threshold to 5.96. The final object ran from the edge of one image stack to another between the first and second night. Our code cannot currently account for this situation in our search. However, we ran the search using only the second and third nights of data to find the object and our code was able to find it at a reduced SNR value of 5.31. This means that when going to a faint enough search threshold we were able to recover all of the known KBOs in the search area.

After submission to the MPC database two more results were matched to objects in the MPC catalogs. Previous observations had not predicted the orbits to fall within the observation fields, but were linked and recalculated by the MPC after submission of our results. These new matches are also noted as previously discovered in Table \ref{object_table}.

\section{Results}

\subsection{Recovery Efficiency} \label{sec:recovery_efficiency}

\begin{figure}
	\centering
    	\includegraphics[width=\linewidth]{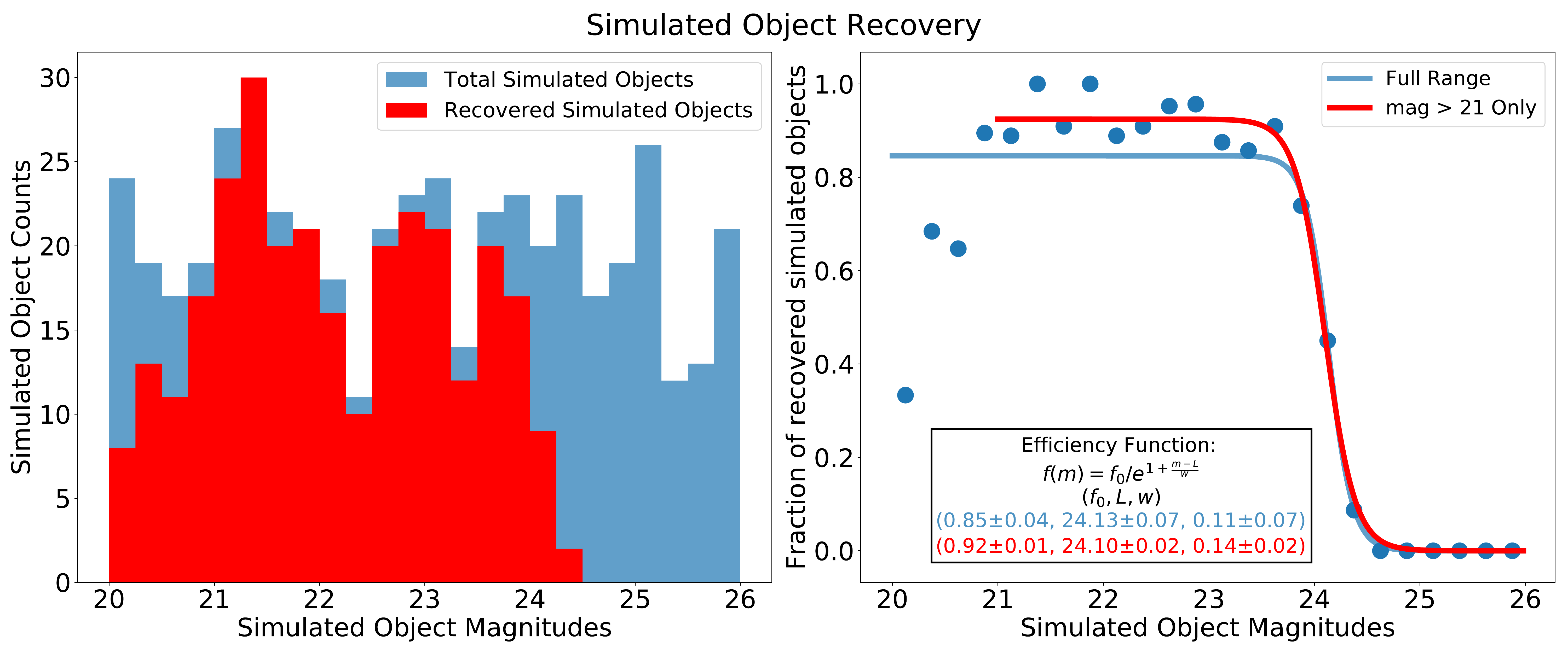}
	\caption{Recovery of simulated objects inserted into a HITS field. Left: Histogram comparing counts of recovered simulated objects to the full set as a function of magnitude. Right: Fraction of recovered simulated objects as a function of magnitude fitted with an efficiency curve for both the full range of simulated objects and for $g > 21$ only.}	\label{simulated_recovery}
\end{figure}

To understand the efficiencies of our search method, we inserted simulated objects with \textit{g} magnitudes between 20-26 into the images from one of the 2015 HITS fields and tested our ability to recover these objects. The objects all had a simple Gaussian PSF that matched our search PSF and the velocity angles and magnitudes were uniformly distributed within the ranges of our search parameters. Figure \ref{simulated_recovery} shows the results in terms of counts on the left and the fraction of the total simulated set on the right. Both plots are a function of DECam \textit{g} magnitude with bin widths of 0.25 magnitudes. For the right plot we fit an efficiency function of the form $f(m) = f_{0} / e^{1 + \frac{m - L}{w}}$, where $f_{0}$ is the efficiency ceiling, $L$ is the 50\% detection probability magnitude, and $w$ is the width in magnitudes of the drop-off in sensitivity.

We plotted two efficiency functions, one for the entire range of magnitudes of the simulated objects and then one at magnitudes $g > 21$. The drop-off at the brighter magnitudes is due to the extra masking we did at a specific count threshold. We will say more about this effect in the discussion in Section \ref{sec:discussion}. On the fainter end we see the limit set by the SNR cutoff at SNR = 10. The detection efficiency ceiling for the full range is $84.6\pm3.7$\% with a 50\% threshold at $g=24.13\pm.07$ and drop-off width of $0.11\pm.07$ mags. When we look at efficiencies at $g > 21$ we get an efficiency ceiling of $92.5\pm1.2$\% and a 50\% detection threshold at the very similar $g=24.10\pm.02$ value and drop-off width of $0.14\pm.02$ mags.

\subsection{Comparison with Existing Models}
While our original searches used only three nights of data, we had additional nights in the HITS data that we used for supplying additional observations when estimating the orbits of the discovered KBOs. This still limited our baseline for the discovered objects to a range from 3-10 days and made accurate estimates of semi-major axis and eccentricity difficult. We have confident measurements on the inclinations and all our inclination estimates for objects in the MPC catalog matched the published inclinations within 1 $\sigma$ of the output values from the \citet{bernstein} orbit fitting code. Therefore, we present a comparison of our inclination distribution to the general KBO population such as that in \citet{brown_inclinations}.

\subsubsection{Inclination Distribution} \label{sec:inclination_comparison}
Our inclination values range from $7^{\circ}$-$44^{\circ}$ where the lower limit comes from the fact that our closest field to the ecliptic is at $-6.4^{\circ}$. Even though our fields are all at moderate latitudes off the ecliptic (as far as $-21.3^{\circ}$), \citet{brown_inclinations} provides a method to compare our results to a predicted distribution by using the inclinations of the objects and the ecliptic latitude of discovery.

\begin{figure}
	\centering
    	\includegraphics[width=\linewidth]{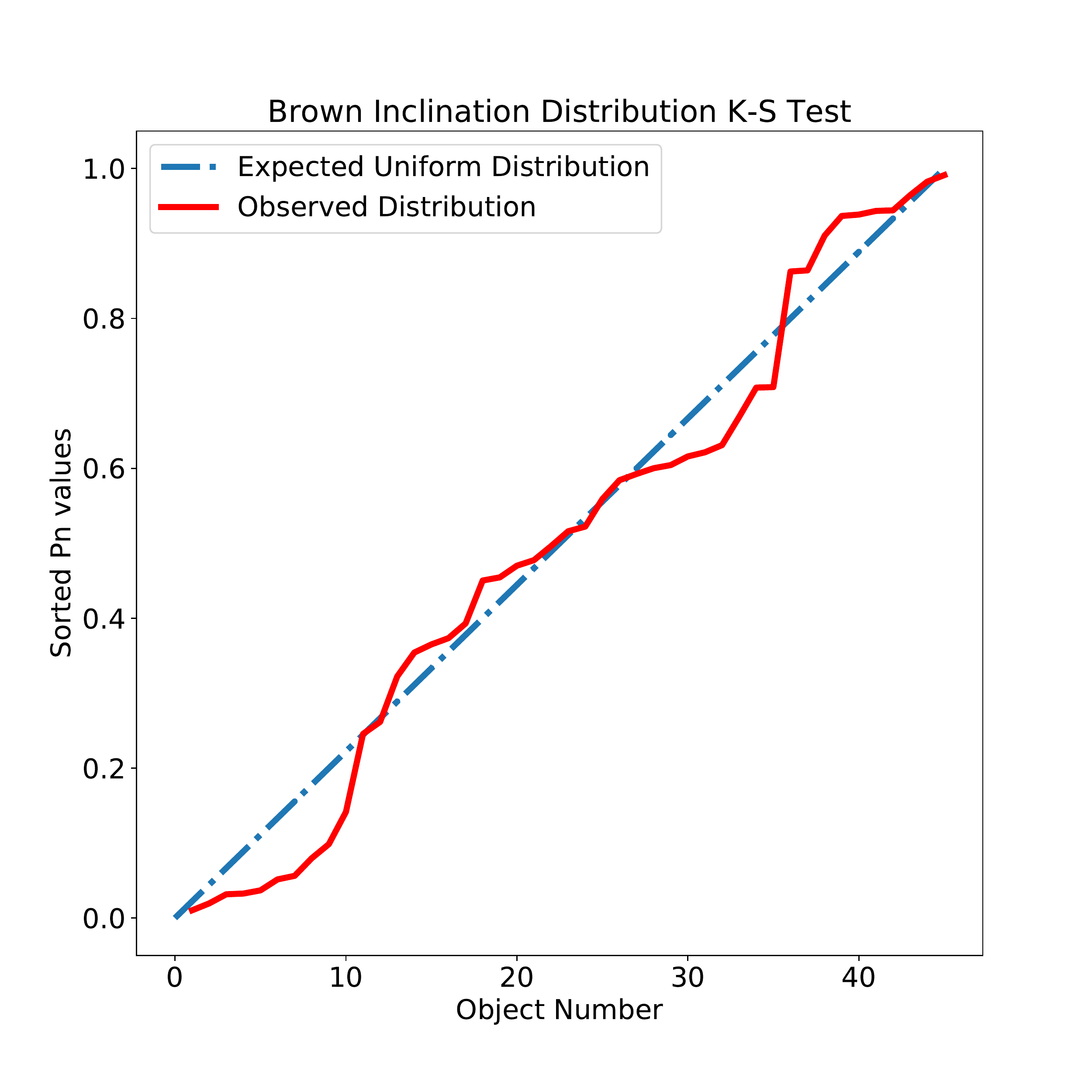}
	\caption{Kuiper variant of K-S Test comparing \citet{brown_inclinations} inclination distribution to our recovered results.}	\label{brown_ks_test}
\end{figure}

\citet{brown_inclinations} estimates an inclination distribution for the full KBO population with a double Gaussian multiplied by $\sin{i}$:
\begin{equation} \label{brown_distribution_equation}
f_t(i) = \sin{i}[a\exp{(\frac{-i^{2}}{2\sigma^{2}_{1}})} + (1-a)\exp{(\frac{-i^{2}}{2\sigma^{2}_{2}})}]
\end{equation}
where $a = 0.89$, $\sigma_{1} = 2.7^\circ$ and $\sigma_{2} = 13.2^\circ$. To compare our results to this distribution we follow the method outlined in Section 3 of \citet{brown_inclinations}. For a given inclination distribution, $f_t$, the probability that an object, $j$, with discovery latitude, $\beta_j$, would have an inclination equal to or below the actual inclination, $i_j$ is given by Equation \ref{discovery_inc_probability}. 
\begin{equation} \label{discovery_inc_probability}
P_j = \int^{i_j}_{\beta_{j}} \frac{f_{t}(i')}{(\sin^{2}i' - \sin^{2}\beta_{j})^{1/2}}di' \times [\int^{\pi/2}_{\beta_{j}}\frac{f_{t}(i')}{(\sin^{2}i' - \sin^{2}\beta_{j})^{1/2}}di']
\end{equation}
The distribution of $P_j$ for the actual KBO population estimated by \citet{brown_inclinations} varies uniformly between 0 and 1. Therefore, if we take as the null hypothesis for our observations that they are an unbiased sample down to our magnitude limits and representative of the distribution of KBO inclinations in the fields we searched we should compare the distributions of $P_j$ for our objects to the uniform distribution. To do this we start by calculating $P_j$ for each of our objects and plot their sorted distribution in Figure \ref{brown_ks_test}. We compare our observed distribution to the inclination distribution of \citet{brown_inclinations} using the Kuiper variant of the Kolmogorov-Smirnov (K-S) test as done in \citet{brown_inclinations} and according to Equation \ref{ks_test_statistic}.
\begin{equation} \label{ks_test_statistic}
D = \max(P_j - j/N)
\end{equation} 
The actual test statistic is $D\sqrt{N}$ where $N$ is the sample size which is $N=45$ in our data set. In order to find the confidence levels for the test, we create $10^5$ sets of N=45 random samples drawn from a uniform distribution and calculate $D\sqrt{N}$ comparing to a uniform distribution. The $1\sigma$ confidence value occurs when the probability of getting higher than a given $D\sqrt{N}$ value is 84.1\%. We find this to be at $D\sqrt{N} = 1.47$.

Finally, we calculate the $P_j$ values using the Monte Carlo methods described in \citet{brown_inclinations}. We first draw $10^5$ inclinations from the \citet{brown_inclinations} distribution and randomly place them along circular orbits. For each of our observed objects, $j$, we then take all of the Monte Carlo objects within $\pm 0.5^\circ$ of the latitude of discovery, $\beta_j$ and construct an empirical inclination distribution. In order to derive $P_j$ for an object, we use this distribution to calculate the probability that an object with the given $\beta_j$ will have an inclination at or below $i_j$. Using this set of $P_j$ values we then perform the K-S test compared to a uniform distribution between 0 and 1. We perform the Monte Carlo simulation 1000 times and use the mean $D\sqrt{N}$ value as our test statistic for comparison. The K-S test comparison for one of the Monte Carlo distributions is shown in Figure \ref{brown_ks_test}. Our mean $D\sqrt{N}$ value after 1000 runs was 1.37, corresponding to a confidence level of 75\% and within the $1\sigma$ level. This means that we cannot reject the hypothesis that our observations come from the \citet{brown_inclinations} distribution and are consistent with this prediction.

\begin{figure}
	\centering
    	\includegraphics[width=\linewidth]{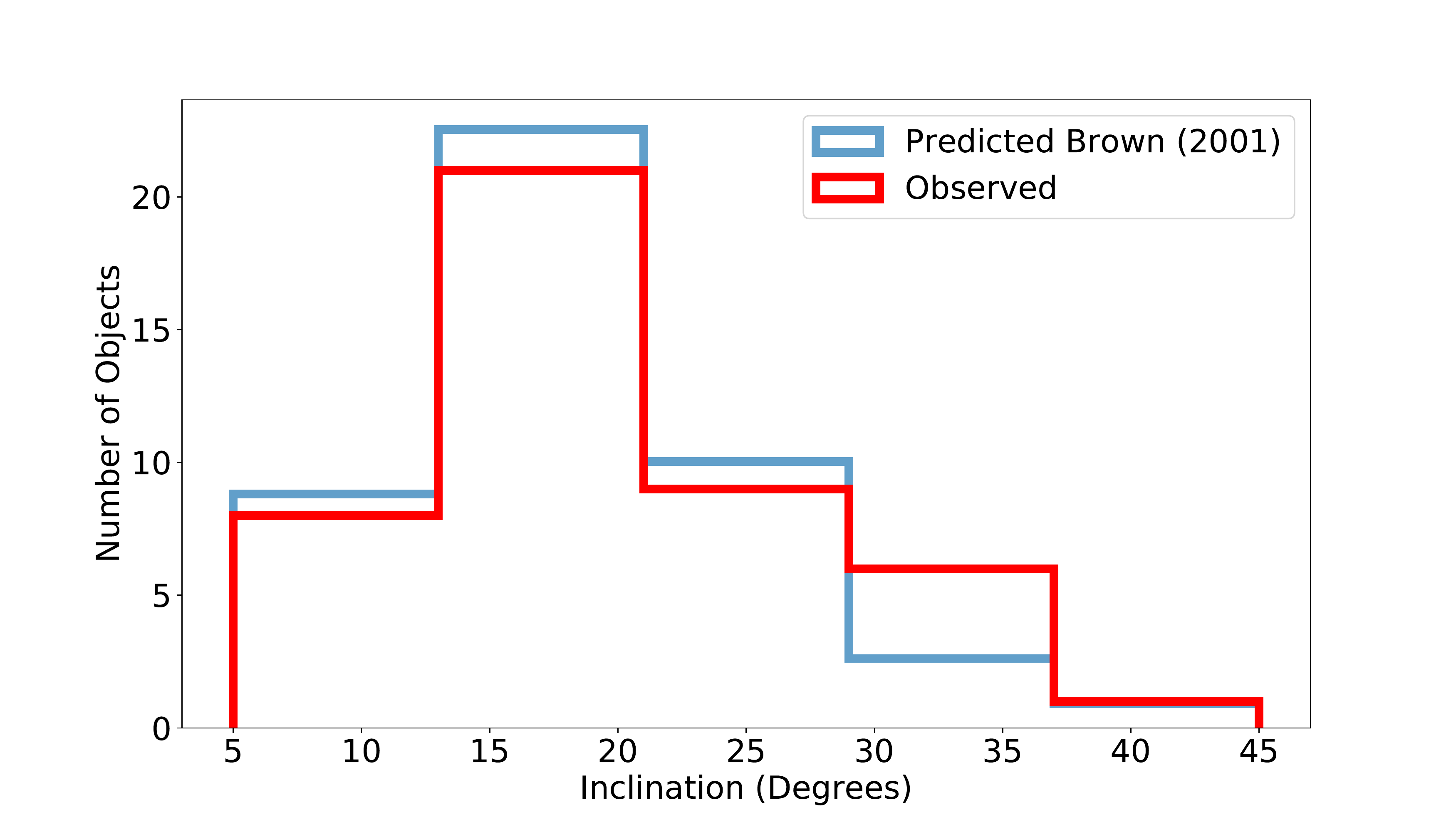}
	\caption{Inclination distributions of detected objects in the HITS fields compared to predicted Brown distribution accounting for the ecliptic latitudes of the HITS observations and normalized to the same number of discovered objects.}	\label{figure:inclination_comparison}
\end{figure}

As a further comparison we did a basic survey simulation to determine the expected distribution of objects we would find in the HITS data. We used the Monte Carlo distribution of inclinations and locations of objects scattered around circular orbits from Figure \ref{brown_ks_test} along with the locations of the HITS fields and approximated the DECam field of view as a circle with $2.2^\circ$ diameter. With this information, we recorded the objects in the Monte Carlo simulation visible through the survey pattern. We then scaled this distribution to the discovery of the same number of objects that we found in the data and plotted them on top of one another in Figure \ref{figure:inclination_comparison}. A $\chi^2$ test on this data gives $\chi^2 = 2.23$ and a p-value of 0.69, meaning we cannot reject the hypothesis that our observed samples come from the \citet{brown_inclinations} inclination distribution function for the general KBO population.

\subsubsection{Magnitude Distribution}
\begin{figure}
	\centering
    	\includegraphics[width=\linewidth]{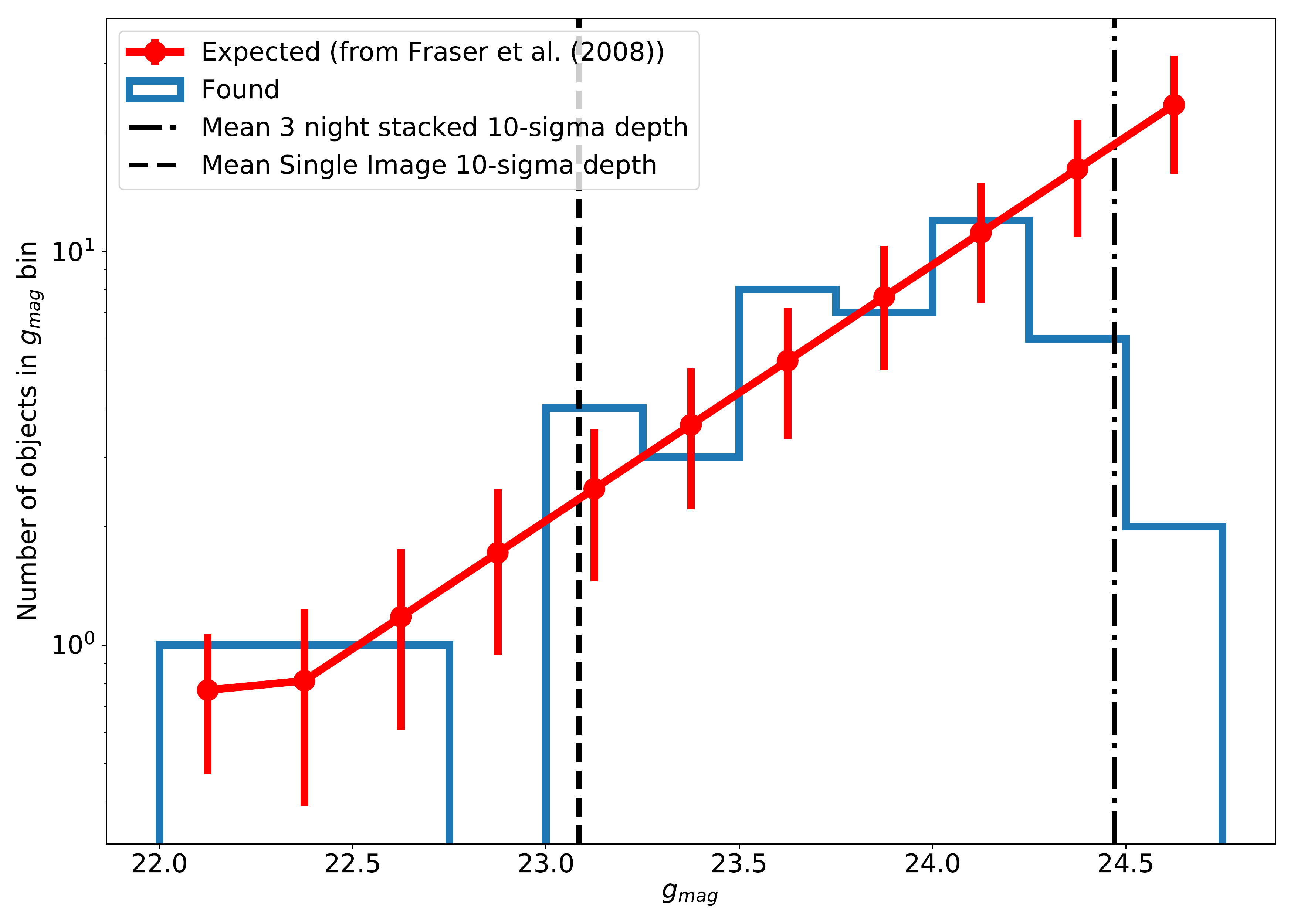}
	\caption{Completeness comparison of KBOs found in HITS survey using KBMOD. Our results are consistent with a complete sample at 24th magnitude compared to the single image depth of 23.1.}	\label{luminosity_comparison}
\end{figure}

We next compare the magnitudes of our discoveries to the KBO magnitude distribution of \citet{fraser_luminosity}. This distribution gives the number of observed objects per square degree as:
\begin{equation} \label{luminosity_distribution}
N_{obs} = 10^{\alpha(m-m_o)}
\end{equation}
where $\alpha = 0.65 \pm 0.05$ and $m_{o} = 23.42 \pm 0.13$ for R magnitudes. Since Equation \ref{luminosity_distribution} is based upon the number of expected objects per square degree at the ecliptic we needed to scale our viewing area appropriately. Satisfied by the work in Section \ref{sec:inclination_comparison}, where we showed our results are consistent with the \citet{brown_inclinations} inclination distribution, we converted this to a latitudinal distribution. We then multiplied the 3 square degree DECam viewing area for each field by the fraction of expected objects at the field's ecliptic latitude compared to the maximum value at the ecliptic. This gave us an effective viewing area of 91.05 square degrees. The next step was converting between the R magnitudes used for Equation \ref{luminosity_distribution} and the \textit{g} magnitudes of our observations. \citet{fraser_luminosity} also had to do magnitude conversions for various datasets and used a KBO $<g'-R>$ color of 0.95 which we use here as well. Putting together the scaling and magnitude offsets we compare our results to the \citet{fraser_luminosity} expected number of objects for our survey fields in Figure \ref{luminosity_comparison}.
The drop-off expected from Section \ref{sec:recovery_efficiency} around 24th magnitude is present and seems to occur after $g=24.25$ after which we fall below 50\% completeness which is consistent with the magnitude where our search drops below 50\% efficiency as shown in Section \ref{sec:recovery_efficiency}. Figure \ref{luminosity_comparison} also shows the 10-sigma threshold we used in our search of the HITS data. A $\chi^2$ test on the data for $g<24.25$ gives $\chi^2 = 10.21$ and a p-value of 0.25, meaning our results are consistent with the \citet{fraser_luminosity} luminosity distribution.

\section{Discussion} \label{sec:discussion}

\subsection{Filtering Analysis}

\begin{figure}
	\centering
    	\includegraphics[width=\linewidth]{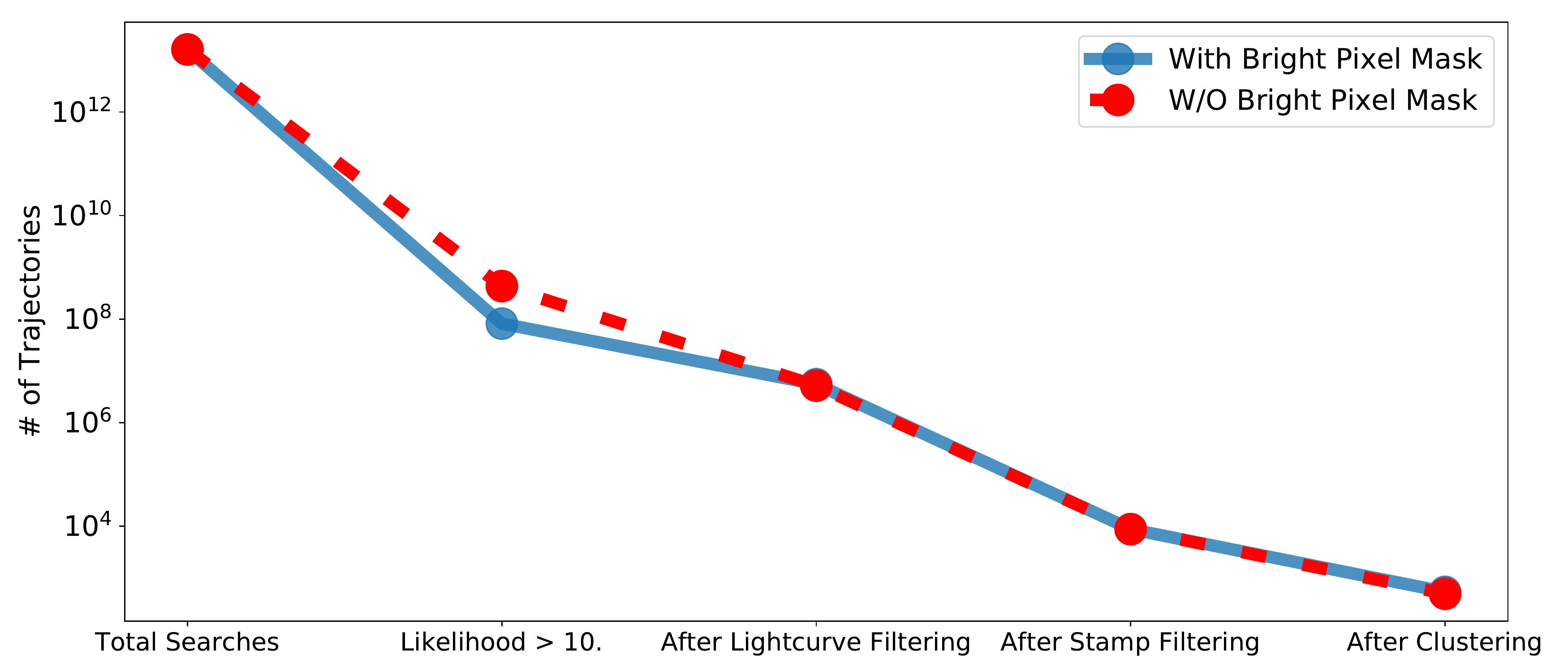}
	\caption{Number of positive trajectories after each step of filtering the search results from a field with simulated objects inserted.}	\label{fig:filtering_effects}
\end{figure}

We used the field with simulated objects from Section \ref{sec:recovery_efficiency} to study the effects of the various stages of our filtering process. The red line in Figure \ref{fig:filtering_effects} shows the results for the searches over the full field in our processing which included an extra masking step that we discuss below. We start with the total number of searches based upon the number of grid steps multiplied by the number of pixels on the focal plane. We only keep those above our detection threshold which is the biggest single reduction in the number of possible results. After that, each step in our filtering process is able to reduce the number of false positives by 1-3 orders of magnitude in the number of total results. The most effective is the postage stamp filtering which uses the moments of the postage stamp image and their resemblance to those of a Gaussian source model.

\subsection{Masking and Threshold Effects} \label{subsec:mask_thresh_effects}

We made a series of decisions in our search based upon our target objects and the use of science images instead of less contaminated difference images. The first choice we made was to include an additional bright pixel mask when creating our $\Psi$ and $\Phi$ images. This mask was in addition to the masking described in Section \ref{sec:search_description}. Any pixel in any image that exceeded 120 counts, corresponding to $g \sim 21$, was masked in our searches. This additional masking covers extended bright halos, that were not covered by our initial footprint masking, and bright fast moving objects which are likely much closer to Earth than our target population. The effects of additional masking on recovery were tested on the simulated object field (Section 4.1). Comparing the red line with this extra masking to the blue line without it in Figure \ref{fig:filtering_effects} shows the major benefit of this masking was a significant decrease in the amount of false positives we had to filter out after our likelihood cut. The amount of positive results after the likelihood cut in Figure \ref{fig:filtering_effects} shows about five times more objects to filter after the threshold cut without the masking. A decrease of $4 \times 10^{8}$ false positives over a single field saves us hours of computation time during our analysis since we processed the results through the lightcurve filter at a rate of 70 seconds for 500,000 objects. Over 50 fields this adds up to days of processing time that we were able to avoid. 

\begin{figure}
	\centering
    	\includegraphics[width=\linewidth]{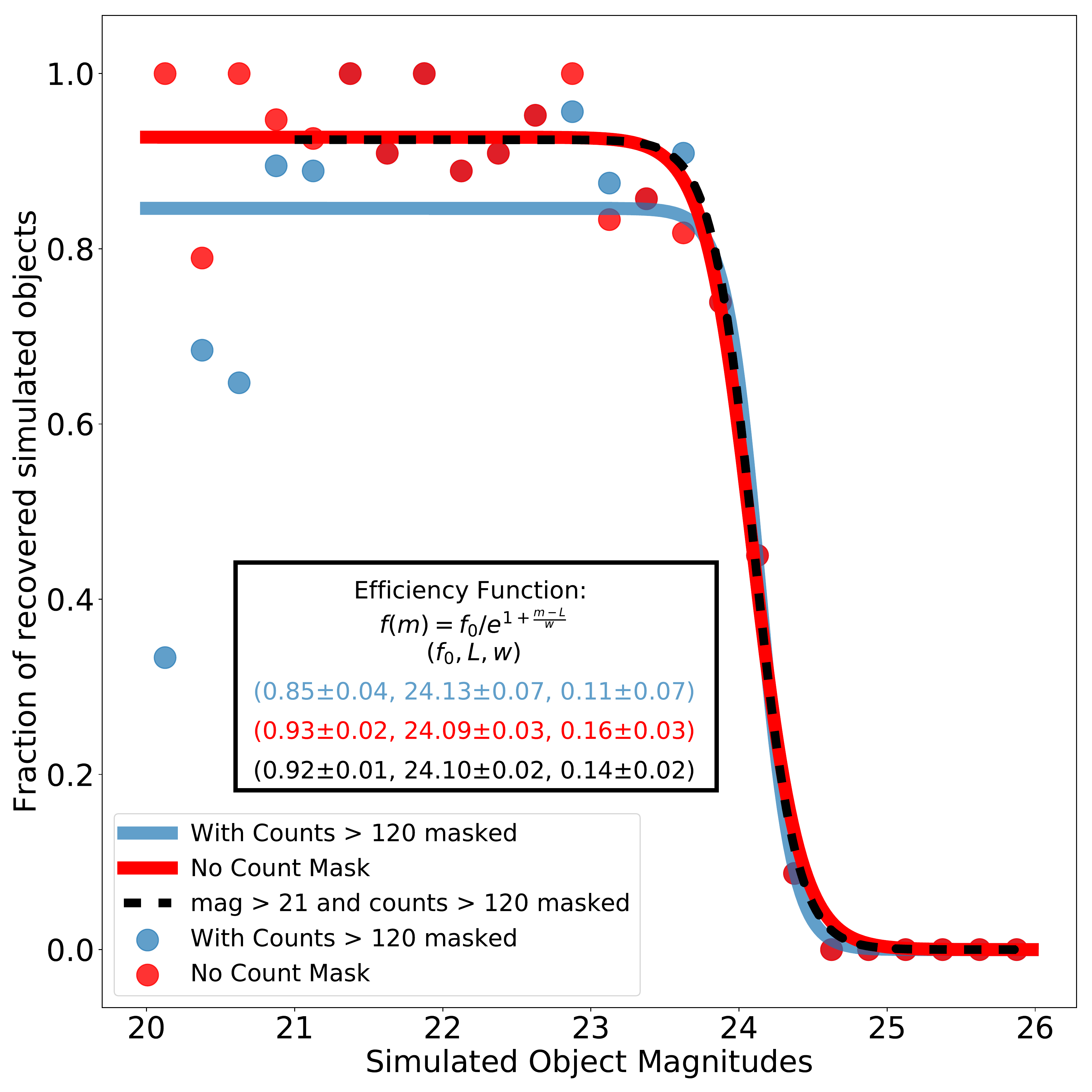}
	\caption{Comparison of efficiency curves run on the same simulated object set with and without the count masking described in Section \ref{subsec:mask_thresh_effects}.  There is almost no effect on the depth of our recoveries.}	\label{fig:simulated_recovery_compare}
\end{figure}

However, we also looked to see what was the price we paid for this extra computational efficiency. Figure \ref{fig:simulated_recovery_compare} compares the efficiency curves of Section \ref{sec:recovery_efficiency} and Figure \ref{simulated_recovery} to the results without the additional bright pixel masking. The recovery efficiency is higher across all magnitudes at $92.7\pm1.5$\%, but when we compare to the recovery at only $g > 21$ with the masking it is very similar to the $92.4\pm1.2$\% of that result. The 50\% efficiency depth is nearly unchanged at $g=24.09\pm.03$ without the masking meaning the bright pixel masking does not affect our overall depth but only the bright end of our recovery at $g < 21$. Using the \citet{fraser_luminosity} curve for the expected number of KBOs for magnitudes $20 < g < 21$ in our survey fields, we calculate that the number of expected objects in this magnitude range to be less than 0.3. We find this to be an acceptable trade off for the computational savings in this particular example. When going to difference imaging or with a different population of brighter expected objects we do not plan to include this extra masking step.

\begin{figure}
	\centering
    	\includegraphics[width=\linewidth]{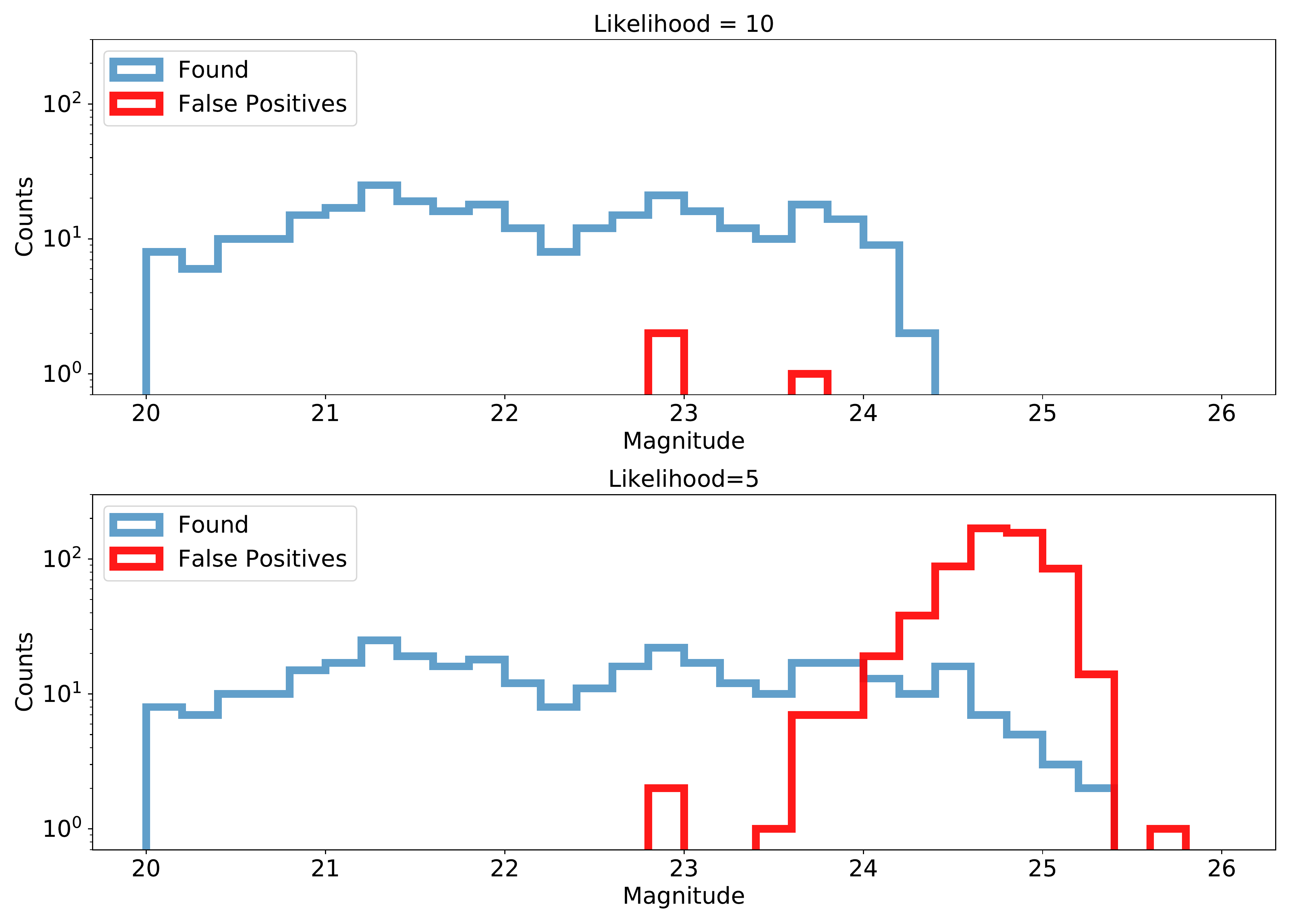}
	\caption{Comparing the false positives at detection thresholds of 10$\sigma$ versus 5$\sigma$ in a field with simulated objects inserted. False positives overwhelm our filtering methods at the lower threshold when using science images.}	\label{fig:false_positive_comparison}
\end{figure}

\begin{figure}
	\centering
    	\includegraphics[width=\linewidth]{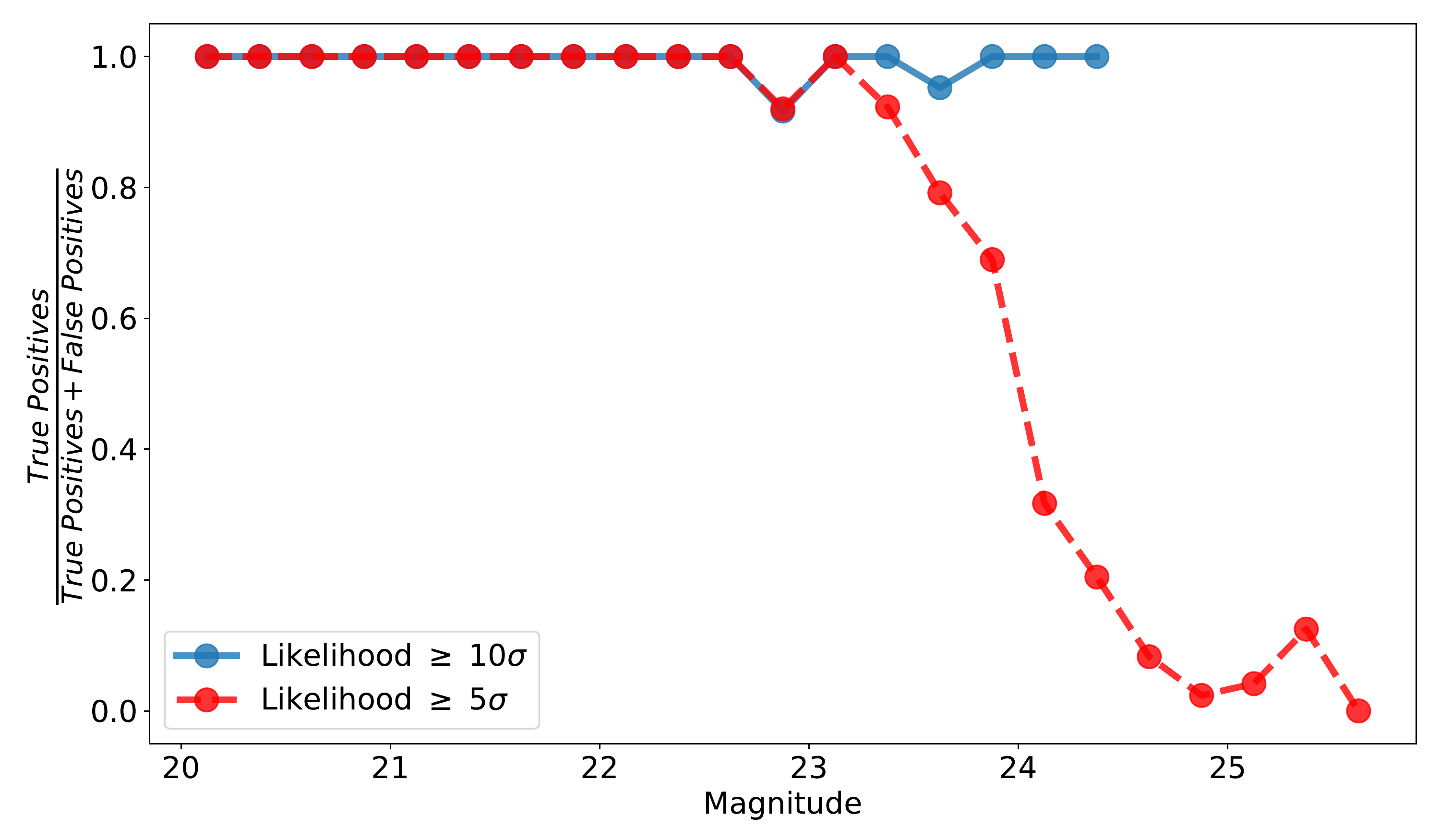}
	\caption{Comparing the precision at detection thresholds of 10$\sigma$ versus 5$\sigma$ in a field with simulated objects inserted. We are very confident in our detections at 10$\sigma$ with science images but would not be if we were to go to 5$\sigma$ detections with the science images.}	\label{fig:precision_comparison}
\end{figure}

Another decision we made was to set our threshold for detection at a limit of 10$\sigma$ instead of a lower, typical catalog level threshold of $5\ \sigma$. This decision was motivated by the use of science images and the extra noise that would be present as this threshold was lowered. We looked at the increase of false positives as we go from 10$\sigma$ to 5$\sigma$ by rerunning the same field as Section \ref{sec:recovery_efficiency} without any simulated objects and the timestamps randomly scrambled so that any detections were false positives. We show histograms comparing the false positives to true detections at the two thresholds in Figure \ref{fig:false_positive_comparison}. We also calculated the precision of our final results where precision is defined as $\frac{True\ Positive}{True\ Positive\ +\ False\ Positive}$ and show the results in Figure \ref{fig:precision_comparison}. While we do detect objects to over a magnitude fainter ($g \sim 25.4$) with the lower likelihood threshold we are overwhelmed by false positives at the fainter magnitudes and even at magnitudes at which we achieve high precision with a higher threshold. At a threshold of 5$\sigma$ we are below 50\% precision at $g = 24.1$ while we are never below 90\% precision when using a 10$\sigma$ threshold. This degradation at a brighter magnitude in the 5$\sigma$ results happens because bright false positives with fewer observations will appear at a higher likelihood than a dim object with more observations (e.g., a greater number of the observations were in a masked area or off the edge of the CCD). Due to this false positive performance when running on science images we used the 10$\sigma$ threshold in this work to show the effectiveness of the algorithm and code while also producing useful results. We discuss our future plans in the next section including what we will do to run the code at a lower detection threshold going forward.

\subsection{Future Work}

There is ongoing work to decrease the detection threshold for sources in order to extend the searches ov
er longer timescales thus allowing us to find fainter and more distant objects. Work with longer baselines and larger stacks of images would also allow us to push this detection limit lower and confidently go beyond the limits of current population studies of KBOs. We plan to use the code with difference images in the future which we hope will reduce the time spent filtering and also remove many of the artifacts from science images that lead to false positives above 5$\sigma$ and also pushed us to add in additional masking procedures in this work. We also plan to move to a deep learning based postage stamp classifier that we hope will perform well at fainter magnitudes and plan to do a comparison versus our existing technique on the difference imaging search. We will also address working with non-linear trajectories and methods to find objects that move from one chip to another during the survey period. Finally, we are also working on enhancements to the GPU algorithm that will allow us to spend less time searching low likelihood trajectories thereby increasing search efficiency.

All of these improvements will help us explore deeper into the Kuiper Belt and enhance our ability to use stacks of images from long baseline surveys such as ZTF or LSST in the future.

\section{Conclusion} \label{sec:conclusion}

In this paper we presented a new algorithm that uses the power of GPU processing to search for slow moving sources across a sequence of images. Our approach is capable of searching over $10^{10}$ candidate moving object trajectories in one minute. Applying these techniques to existing data 
we discovered 39 new KBOs and the recovered 6 KBOs already present in the Minor Planet Center catalogs. Finally, we used the results of our search to compare to the \citet{brown_inclinations} Kuiper Belt inclination distribution and the \citet{fraser_luminosity} magnitude distribution. We found both of the models to be consistent with observed results indicating that the recovered sample matches overall published characteristics of the KBO population. Combined with the high rates of detection efficiency recovered in tests, this indicates that our software provides a nearly complete recovery of an unbiased sample of moving objects down to the detection limit.

Our software, Kernel Based Moving Object Detection (KBMOD) is available to the public at \url{https://github.com/dirac-institute/kbmod}. This includes the GPU searching code as well as python analysis code we have used to process the search results. Development is continuing and can be followed at the Github repository hosting the code.

\acknowledgements
We would like to thank the anonymous referee for very helpful feedback that improved this paper. We also wish to thank Jim Bosch for valuable insights when we were starting this project. AJC and JBK acknowledge partial support from NSF awards AST-1409547 and OAC-1739419. AJC, JBK, PW  acknowledge support from the DIRAC Institute in the Department of Astronomy at the University of Washington. The DIRAC Institute is supported through generous gifts from the Charles and Lisa Simonyi Fund for Arts and Sciences, and the Washington Research Foundation.



\software{KBMOD \citep{whidden+2018}, LSST DM Stack \citep{lsst-dm}, astropy \citep{astropy}, scikit-image \citep{skimage}, numpy \citep{numpy}, CUDA \citep{cuda}, scikit-learn \citep{scikit}, pandas \citep{pandas}, matplotlib \citep{matplotlib}}

\bibliography{kbmod}{}

\begin{thebibliography}{}
\expandafter\ifx\csname natexlab\endcsname\relax\def\natexlab#1{#1}\fi

\bibitem[{{Allen} {et~al.}(2001){Allen}, {Bernstein}, \&
  {Malhotra}}]{Allen+2001}
{Allen}, R.~L., {Bernstein}, G.~M., \& {Malhotra}, R. 2001, \apjl, 549, L241

\bibitem[{{Astropy Collaboration} {et~al.}(2013){Astropy Collaboration},
  {Robitaille}, {Tollerud}, {Greenfield}, {Droettboom}, {Bray}, {Aldcroft},
  {Davis}, {Ginsburg}, {Price-Whelan}, {Kerzendorf}, {Conley}, {Crighton},
  {Barbary}, {Muna}, {Ferguson}, {Grollier}, {Parikh}, {Nair}, {Unther},
  {Deil}, {Woillez}, {Conseil}, {Kramer}, {Turner}, {Singer}, {Fox}, {Weaver},
  {Zabalza}, {Edwards}, {Azalee Bostroem}, {Burke}, {Casey}, {Crawford},
  {Dencheva}, {Ely}, {Jenness}, {Labrie}, {Lim}, {Pierfederici}, {Pontzen},
  {Ptak}, {Refsdal}, {Servillat}, \& {Streicher}}]{astropy}
{Astropy Collaboration}, {Robitaille}, T.~P., {Tollerud}, E.~J., {et~al.} 2013,
  \aap, 558, A33

\bibitem[{{Bernstein} \& {Khushalani}(2000)}]{bernstein}
{Bernstein}, G., \& {Khushalani}, B. 2000, \aj, 120, 3323

\bibitem[{{Bernstein} {et~al.}(2004){Bernstein}, {Trilling}, {Allen}, {Brown},
  {Holman}, \& {Malhotra}}]{Bernstein+2004}
{Bernstein}, G.~M., {Trilling}, D.~E., {Allen}, R.~L., {et~al.} 2004, \aj, 128,
  1364

\bibitem[{{Bosch}(2015)}]{bosch}
{Bosch}, J. 2015, Algorithms for Detection and Coaddition.
  \url{https://github.com/lsst-dm/algorithm-docs/blob/master/2015-07-det\%2Bcoadd-slides/slides.ipynb}

\bibitem[{Bosch {et~al.}(2017)Bosch, Armstrong, Bickerton, Furusawa, Ikeda,
  Koike, Lupton, Mineo, Price, Takata, Tanaka, Yasuda, AlSayyad, Becker,
  Coulton, Coupon, Garmilla, Huang, Krughoff, Lang, Leauthaud, Lim, Lust,
  MacArthur, Mandelbaum, Miyatake, Miyazaki, Murata, More, Okura, Owen,
  Swinbank, Strauss, Yamada, \& Yamanoi}]{Bosch+2017}
Bosch, J., Armstrong, R., Bickerton, S., {et~al.} 2017, arXiv.org, 502

\bibitem[{{Brown}(2001)}]{brown_inclinations}
{Brown}, M.~E. 2001, \aj, 121, 2804

\bibitem[{{Denneau} {et~al.}(2013){Denneau}, {Jedicke}, {Grav}, {Granvik},
  {Kubica}, {Milani}, {Vere{\v s}}, {Wainscoat}, {Chang}, {Pierfederici},
  {Kaiser}, {Chambers}, {Heasley}, {Magnier}, {Price}, {Myers}, {Kleyna},
  {Hsieh}, {Farnocchia}, {Waters}, {Sweeney}, {Green}, {Bolin}, {Burgett},
  {Morgan}, {Tonry}, {Hodapp}, {Chastel}, {Chesley}, {Fitzsimmons}, {Holman},
  {Spahr}, {Tholen}, {Williams}, {Abe}, {Armstrong}, {Bressi}, {Holmes},
  {Lister}, {McMillan}, {Micheli}, {Ryan}, {Ryan}, \& {Scotti}}]{denneau+2013}
{Denneau}, L., {Jedicke}, R., {Grav}, T., {et~al.} 2013, \pasp, 125, 357

\bibitem[{Ester {et~al.}(1996)Ester, Kriegel, Sander, \& Xu}]{dbscan}
Ester, M., Kriegel, H.-P., Sander, J., \& Xu, X. 1996, in Proceedings of the
  Second International Conference on Knowledge Discovery and Data Mining,
  KDD'96 (AAAI Press), 226--231

\bibitem[{Flaugher {et~al.}(2015)Flaugher, Diehl, Honscheid, Abbott, Alvarez,
  Angstadt, Annis, Antonik, Ballester, Beaufore, Bernstein, Bernstein, Bigelow,
  Bonati, Boprie, Brooks, Buckley-Geer, Campa, Cardiel-Sas, Castander,
  Castilla, Cease, Cela-Ruiz, Chappa, Chi, Cooper, da~Costa, Dede, Derylo,
  DePoy, de~Vicente, Doel, Drlica-Wagner, Eiting, Elliott, Emes, Estrada, Neto,
  Finley, Flores, Frieman, Gerdes, Gladders, Gregory, Gutierrez, Hao, Holland,
  Holm, Huffman, Jackson, James, Jonas, Karcher, Karliner, Kent, Kessler,
  Kozlovsky, Kron, Kubik, Kuehn, Kuhlmann, Kuk, Lahav, Lathrop, Lee, Levi,
  Lewis, Li, Mandrichenko, Marshall, Martinez, Merritt, Miquel, Muñoz,
  Neilsen, Nichol, Nord, Ogando, Olsen, Palaio, Patton, Peoples, Plazas, Rauch,
  Reil, Rheault, Roe, Rogers, Roodman, Sanchez, Scarpine, Schindler, Schmidt,
  Schmitt, Schubnell, Schultz, Schurter, Scott, Serrano, Shaw, Smith,
  Soares-Santos, Stefanik, Stuermer, Suchyta, Sypniewski, Tarle, Thaler, Tighe,
  Tran, Tucker, Walker, Wang, Watson, Weaverdyck, Wester, Woods, Yanny, \&
  Collaboration}]{Flaugher_DECam_Instrument}
Flaugher, B., Diehl, H.~T., Honscheid, K., {et~al.} 2015, The Astronomical
  Journal, 150, 150

\bibitem[{{F{\"o}rster} {et~al.}(2016){F{\"o}rster}, {Maureira}, {San
  Mart{\'i}n}, {Hamuy}, {Mart{\'i}nez}, {Huijse}, {Cabrera}, {Galbany}, {de
  Jaeger}, {Gonz{\'a}lez-Gait{\'a}n}, {Anderson}, {Kunkarayakti}, {Pignata},
  {Bufano}, {Litt{\'i}n}, {Olivares}, {Medina}, {Smith}, {Vivas},
  {Est{\'e}vez}, {Mu{\~n}oz}, \& {Vera}}]{HITS}
{F{\"o}rster}, F., {Maureira}, J.~C., {San Mart{\'i}n}, J., {et~al.} 2016,
  \apj, 832, 155

\bibitem[{{Fraser} {et~al.}(2008){Fraser}, {Kavelaars}, {Holman}, {Pritchet},
  {Gladman}, {Grav}, {Jones}, {MacWilliams}, \& {Petit}}]{fraser_luminosity}
{Fraser}, W.~C., {Kavelaars}, J.~J., {Holman}, M.~J., {et~al.} 2008, \icarus,
  195, 827

\bibitem[{{Gladman} \& {Kavelaars}(1997)}]{Gladman+1997}
{Gladman}, B., \& {Kavelaars}, J.~J. 1997, \aap, 317, L35

\bibitem[{{Heinze} {et~al.}(2015){Heinze}, {Metchev}, \&
  {Trollo}}]{Heinze+2015}
{Heinze}, A.~N., {Metchev}, S., \& {Trollo}, J. 2015, \aj, 150, 125

\bibitem[{Hunter(2007)}]{matplotlib}
Hunter, J.~D. 2007, Computing In Science \& Engineering, 9, 90

\bibitem[{Johnston \& Krishnamurthy(2002)}]{johnston+2002}
Johnston, L.~A., \& Krishnamurthy, V. 2002, IEEE Transactions on Aerospace and
  Electronic Systems, 38, 228

\bibitem[{{Juri{\'c}} {et~al.}(2015){Juri{\'c}}, {Kantor}, {Lim}, {Lupton},
  {Dubois-Felsmann}, {Jenness}, {Axelrod}, {Aleksi{\'c}}, {Allsman},
  {AlSayyad}, {Alt}, {Armstrong}, {Basney}, {Becker}, {Becla}, {Bickerton},
  {Biswas}, {Bosch}, {Boutigny}, {Carrasco Kind}, {Ciardi}, {Connolly},
  {Daniel}, {Daues}, {Economou}, {Chiang}, {Fausti}, {Fisher-Levine},
  {Freemon}, {Gee}, {Gris}, {Hernandez}, {Hoblitt}, {Ivezi{\'c}}, {Jammes},
  {Jevremovi{\'c}}, {Jones}, {Bryce Kalmbach}, {Kasliwal}, {Krughoff}, {Lang},
  {Lurie}, {Lust}, {Mullally}, {MacArthur}, {Melchior}, {Moeyens}, {Nidever},
  {Owen}, {Parejko}, {Peterson}, {Petravick}, {Pietrowicz}, {Price}, {Reiss},
  {Shaw}, {Sick}, {Slater}, {Strauss}, {Sullivan}, {Swinbank}, {Van Dyk},
  {Vuj{\v c}i{\'c}}, {Withers}, {Yoachim}, \& {LSST Project}}]{lsst-dm}
{Juri{\'c}}, M., {Kantor}, J., {Lim}, K., {et~al.} 2015, ArXiv e-prints,
  arXiv:1512.07914

\bibitem[{Kaiser(2004)}]{kaiser}
Kaiser, N. 2004, Addition of Images with Varying Seeing.
  \url{http://spider.ipac.caltech.edu/staff/fmasci/home/astro_refs/PanStars_Coadder.pdf}

\bibitem[{Kalman(1960)}]{kalman}
Kalman, R.~E. 1960, Journal of Basic Engineering, 82, 35

\bibitem[{{Kubica} {et~al.}(2007){Kubica}, {Denneau}, {Grav}, {Heasley},
  {Jedicke}, {Masiero}, {Milani}, {Moore}, {Tholen}, \&
  {Wainscoat}}]{kubica+2007}
{Kubica}, J., {Denneau}, L., {Grav}, T., {et~al.} 2007, \icarus, 189, 151

\bibitem[{{Lang} {et~al.}(2009){Lang}, {Hogg}, {Jester}, \& {Rix}}]{Lang+2009}
{Lang}, D., {Hogg}, D.~W., {Jester}, S., \& {Rix}, H.-W. 2009, \aj, 137, 4400

\bibitem[{McKinney(2010)}]{pandas}
McKinney, W. 2010, in Proceedings of the 9th Python in Science Conference, ed.
  S.~van~der Walt \& J.~Millman, 51 -- 56

\bibitem[{Nickolls {et~al.}(2008)Nickolls, Buck, Garland, \& Skadron}]{cuda}
Nickolls, J., Buck, I., Garland, M., \& Skadron, K. 2008, Queue, 6, 40

\bibitem[{Oliphant(2006)}]{numpy}
Oliphant, T.~E. 2006, A guide to NumPy (USA: Trelgol Publishing)

\bibitem[{{Pedregosa} {et~al.}(2012){Pedregosa}, {Varoquaux}, {Gramfort},
  {Michel}, {Thirion}, {Grisel}, {Blondel}, {M{\"u}ller}, {Nothman}, {Louppe},
  {Prettenhofer}, {Weiss}, {Dubourg}, {Vanderplas}, {Passos}, {Cournapeau},
  {Brucher}, {Perrot}, \& {Duchesnay}}]{scikit}
{Pedregosa}, F., {Varoquaux}, G., {Gramfort}, A., {et~al.} 2012, ArXiv
  e-prints, arXiv:1201.0490

\bibitem[{Reed {et~al.}(1988)Reed, Gagliardi, \& Stotts}]{reed+1988}
Reed, I.~S., Gagliardi, R.~M., \& Stotts, L.~B. 1988, IEEE Transactions on
  Aerospace and Electronic Systems, 24, 327

\bibitem[{{Rozovskii} \& {Petrov}(1999)}]{rozovskii+1999}
{Rozovskii}, B.~L., \& {Petrov}, A. 1999, in \procspie, Vol. 3809, Signal and
  Data Processing of Small Targets 1999, ed. O.~E. {Drummond}, 152--163

\bibitem[{{Shao} {et~al.}(2014){Shao}, {Nemati}, {Zhai}, {Turyshev}, {Sandhu},
  {Hallinan}, \& {Harding}}]{Shao+2014}
{Shao}, M., {Nemati}, B., {Zhai}, C., {et~al.} 2014, \apj, 782, 1

\bibitem[{{Szalay} {et~al.}(1999){Szalay}, {Connolly}, \&
  {Szokoly}}]{szalay+1999}
{Szalay}, A.~S., {Connolly}, A.~J., \& {Szokoly}, G.~P. 1999, \aj, 117, 68

\bibitem[{van~der Walt {et~al.}(2014)van~der Walt, {S}ch\"onberger,
  {Nunez-Iglesias}, {B}oulogne, {W}arner, {Y}ager, {G}ouillart, {Y}u, \& the
  scikit-image contributors}]{skimage}
van~der Walt, S., {S}ch\"onberger, J.~L., {Nunez-Iglesias}, J., {et~al.} 2014,
  PeerJ, 2, e453

\bibitem[{Whidden {et~al.}(2018)Whidden, Kalmbach, \&
  Smotherman}]{whidden+2018}
Whidden, P., Kalmbach, J.~B., \& Smotherman, H. 2018, dirac-institute/kbmod:
  Paper Release, doi:10.5281/zenodo.1342298

\end{thebibliography}



\appendix
    
    
\section{Table of Detected Objects} \label{sec:object_table}

\startlongtable
\begin{deluxetable}{cccccc}
\tablecaption{Detected KBOs in HiTS field \label{object_table} with estimated orbit properties from HiTS data}
\startdata
MPC designation & Semi-Major Axis (AU) & Eccentricity & Inclination & g\_mag & New Discovery?         \\
2015 DZ248      & 67.33                   & 0.42        & 17.60        & 23.68  & Yes              \\
2015 DT248     & 21.38                   & 0.36        & 22.21       & 24.14  & Yes              \\
2015 DT249     & 43.67                   & 0.20         & 25.76       & 23.74  & Yes              \\
2015 DY248   & 48.02                   & 0.28        & 14.72       & 24.19  & Yes              \\
2015 DX248    & 37.61                   & 0.02        & 24.67       & 24.24  & Yes              \\
2015 DV248   & 42.32                   & 0.67        & 27.67       & 23.94  & Yes              \\
2015 DW248   & 48.52                   & 0.62        & 31.42       & 23.02  & Yes              \\
2015 DU248      & 44.24                   & 0.02        & 14.44       & 23.99  & Yes              \\
2015 DQ248  & 63.15                   & 0.45        & 33.13       & 24.05  & Yes              \\
2015 DR248   & 41.61                   & 0.31        & 15.81       & 23.95  & Yes              \\
2015 DS248    & 42.67                   & 0.78        & 23.02       & 24.00     & Yes              \\
2015 DO248 & 46.90                    & 0.27        & 22.65       & 24.08  & Yes              \\
2015 DP248  & 48.97                   & 0.70         & 21.73       & 24.15  & Yes              \\
2015 DO249  & 48.99                   & 0.29        & 19.88       & 24.35  & Yes              \\
2015 DP249  & 49.49                   & 0.43        & 17.06       & 24.28  & Yes              \\
2015 DZ249  & 45.05                   & 0.25        & 14.27       & 24.30   & Yes              \\
2015 DY249 & 38.14                   & 0.03        & 25.72       & 24.41  & Yes              \\
2015 DN249  & 38.30                    & 0.03        & 17.08       & 24.34  & Yes              \\
2015 DM249  & 44.39                   & 0.14        & 12.05       & 23.74  & Yes              \\
2015 DX249    & 47.08                   & 0.24        & 30.04       & 23.22  & Yes              \\
2015 DL249   & 44.18                   & 0.27        & 13.83       & 24.23  & Yes              \\
2015 DK249   & 44.29                   & 0.02        & 7.03        & 24.05  & Yes              \\
2015 DH249   & 46.85                   & 0.02        & 20.71       & 24.63  & Yes              \\
2015 DJ249 & 42.81                   & 0.19        & 12.99       & 23.44  & Yes              \\
2015 DW249 & 38.83                   & 0.19        & 15.11       & 23.68  & Yes              \\
2015 DC249    & 41.67                   & 0.16        & 20.77       & 23.62  & Yes              \\
2015 DD249    & 65.74                   & 0.32        & 19.25       & 23.65  & Yes              \\
2015 DB249  & 40.31                   & 0.27        & 24.58       & 23.56  & Yes              \\
2015 DU249  & 59.50                    & 0.21        & 18.87       & 24.72  & Yes              \\
2015 DA249  & 38.40                    & 0.15        & 19.97       & 23.61  & Yes              \\
2015 DF249  & 45.35                   & 0.44        & 12.99       & 24.08  & Yes              \\
2014 BV64 & 50.20                    & 0.31        & 15.43       & 22.10   & No  \\
2015 DE249  & 41.66                   & 0.35        & 20.29       & 23.35  & Yes              \\
2015 DV249    & 54.12                   & 0.15        & 11.79       & 24.00     & Yes              \\
2015 BF519   & 45.12                   & 0.39        & 18.42       & 23.00     & No              \\
2015 DG249    & 64.80                    & 0.33        & 15.44       & 24.19  & Yes              \\
2011 CX119   & 47.70                    & 0.46        & 29.47       & 23.19  & No \\
2015 DA250    & 45.12                   & 0.29        & 34.11       & 23.89  & Yes              \\
2015 DB250   & 44.13                   & 0.27        & 10.99       & 23.80   & Yes              \\
2014 XW40  & 58.03                   & 0.32        & 16.43       & 23.39  & No              \\
2015 DQ249  & 28.07                   & 0.44        & 43.91       & 23.93  & Yes              \\
2015 DR249    & 41.10                    & 0.02        & 35.13       & 23.81  & Yes              \\
2014 XP40   & 78.02                   & 0.62        & 11.53       & 22.33  & No  \\
2013 FZ27  & 57.12                   & 0.25        & 13.54       & 22.73  & No  \\
2015 DS249  & 31.08                   & 0.29        & 11.38       & 24.37  & Yes             
\enddata
\end{deluxetable}

\end{document}